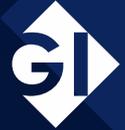

**EXPERT ASSESSMENT**

# THE SYSTEMIC ENVIRONMENTAL RISKS OF ARTFICIAL INTELLIGENCE

DECEMBER 2025

**EXPERT ASSESSMENT**

# THE SYSTEMIC ENVIRONMENTAL RISKS OF ARTFICIAL INTELLIGENCE

DECEMBER 2025

# Legal information

**Publisher:**
Gesellschaft für Informatik e.V. (GI)
Weydingerstraße 14–16
10178 Berlin
Deutschland
Tel: +49 30 240 09-805
E-Mail: berlin@gi.de
Internet: https://gi.de

**Authors:**
Julian Schön, GI
Lena Hoffmann, GI
Nikolas Becker, GI

**As of:**
December 2025

**Acknowledgments:**
This report was prepared as part of the project 'Systemic and Existential Risks of Artificial Intelligence' (SERI) funded by the Federal Ministry of Research, Technology, and Space (Grant No.: 16IS23075). The Institute for Technology Assessment and Systems Analysis acted as principal. The report was commissioned and supervised by Lucas Staab and Carsten Orwat.

SERI investigates the systemic and existential risks to society arising from the development and use of artificial intelligence with the aim of identifying options for action in politics, research, and development, thereby contributing to the mitigation of these risks.

The authors would like to thank Frank Oliver Glöckner, Berk Calli, Namrata Bist, Zane Swanson, and those who wished to remain anonymous for providing their insights for the case studies. Additionally, we would like to thank Verena Majunke, Raghavendra Selvan, Leona Harting, and Jake Farell for providing invaluable feedback.

AI has been used in the creation of this manuscript in the following way: linguistic improvement (Grammarly, Claude, Gemini), translation (DeepL), transcription (Vibe), document summarization and querying (NotebookLM).

With funding from the:

Federal Ministry of Research, Technology and Space

# Contents

# Tables

# Abbreviations

| | |
|---|---|
| 3TG | Tin, Tantalum, Tungsten, Gold |
| AI | Artificial Intelligence |
| ASICs | Application-Specific Integrated Circuits |
| $CO_2$-eq | $CO_2$-equivalent |
| CPU | Central Processing Unit |
| CSR | Corporate Social Responsibility |
| DL | Deep Learning |
| DRAM | Dynamic Random Access Memory |
| E-Waste | Electronic Waste |
| EOL | End-of-Life |
| EU | European Union |
| Fab | Fabrication Facility |
| FLOP | Floating Point Operation |
| FPGAs | Field Programmable Gate Arrays |
| GDP | Gross Domestic Product |
| GHG | Greenhouse Gas |
| GPU | Graphics Processing Unit |
| GWP | Global Warming Potential |
| HBM | High-Bandwidth Memory |
| HIC | High-Income Country |
| ICT | Information and Communication Technology |
| LCA | Life Cycle Assessment |
| LLM | Large Language Model |
| LMIC | Low- to Middle-Income Country |
| ML | Machine Learning |
| MoE | Mixture-of-Experts |
| NAS | Neural Architecture Search |
| NLP | Natural Language Processing |
| PFAS | Per- and Polyfluoroalkyl Substances |
| PUE | Power Usage Effectiveness |
| REE | Rare Earth Element |
| SSD | Solid State Drive |
| TSMC | Taiwan Semiconductor Manufacturing Company |
| TPU | Tensor Processing Unit |
| UPW | Ultra-Pure Water |
| WUE | Water Usage Effectiveness |

# Executive Summary

Artificial intelligence (AI) is often presented as a key tool for addressing societal challenges, such as climate change. At the same time, AI's environmental footprint is expanding increasingly. This report describes the systemic environmental risks of artificial intelligence, in particular, moving beyond direct impacts such as energy and water usage. Systemic environmental risks of AI are emergent, cross-sector harms to climate, biodiversity, freshwater, and broader socioecological systems that arise primarily from AI's integration into social, economic, and physical infrastructures, rather than its direct resource use, and that propagate through feedbacks, yielding nonlinear, inequitable, and potentially irreversible impacts.

While these risks are emergent and quantification is uncertain, this report aims to provide an overview of systemic environmental risks. Drawing on a narrative literature review, we propose a three-level framework that operationalizes systemic risk analysis. The framework identifies the structural conditions that shape AI development, the risk amplification mechanisms that propagate environmental harm, and the impacts that manifest as observable ecological and social consequences. We illustrate the framework in expert-interview-based case studies across agriculture and biodiversity, oil and gas, and waste management.

AI's systemic environmental risks are structural. Our analysis finds they are enabled by power concentration, limited governance, market-driven imperatives, and opacity. These conditions support risk mechanisms such as rebound effects and path dependencies. The resulting impacts include material harms such as resource depletion and toxicity, the erosion of local and Indigenous knowledge, and the weakening of socio-ecological resilience. The distribution of burdens and benefits is inequitable. Environmental harm disproportionately falls on marginalized communities, while benefits accumulate in a few resource-rich actors.

We recommend a precautionary governance framework that treats AI as planetary-scale infrastructure. We propose four policy interventions to address these dynamics:

- Mandated and standardized life cycle and systemic risk assessments, with proportionality tests that weigh social benefits against environmental and justice costs, should be prerequisites for model release, large-scale deployment, and major infrastructure expansion.
- Replace self-reporting with transparency and independent verification across data provenance, models, and direct and systemic environmental impacts, with public reporting of standardized metrics.
- Establish a precautionary global governance framework that treats AI as infrastructure with planetary-scale consequences and aligns development with planetary boundaries.
- Democratize data, models, and infrastructure, and ensure that development and deployment are participatory.

We conclude that a paradigm shift is needed to address the systemic environmental risks of AI adequately. Rather than focusing on isolated measures such as algorithmic efficiency or voluntary self-reporting, this requires a holistic approach that recognizes AI as a global infrastructure. As such, aligning its development and use with finite Earth systems should not be considered an aspiration but a near-term necessity.

# 1. Introduction

Artificial Intelligence, or AI, is increasingly becoming ubiquitous. It is transforming industries such as finance, advertising, commerce, mobility, and healthcare, and is widely seen as an enabling technology with the potential to transform productivity, innovation, and public services across sectors. Consequently, AI has been promoted as a tool to tackle climate change by actors including international bodies (Clutton-Brock et al., 2021; IEA, 2025; Obringer et al., 2024; UNFCCC, 2025; World Economic Forum, 2021), major technology corporations (Amazon, 2025; Dannouni et al., 2023; Google, 2025; Meta, 2025; Microsoft, 2025; NVIDIA, 2025), and academic publications (Rolnick et al., 2022; Stern et al., 2025; Vinuesa et al., 2020). Simultaneously, AI's large and rapidly growing environmental footprint itself has become a topic of attention. Training costs for frontier models have grown about 2.4× per year since 2016, suggesting rising electricity demand and a larger carbon footprint unless grids decarbonize (Cottier et al., 2025). If this trend continues, the amortized training cost will exceed one billion dollars by 2027 (Cottier et al., 2025). Further studies indicate that the computations required for Deep Learning (DL) grew by a factor of 300,000 between 2012 and 2018 (Schwartz et al., 2020), with a single training run of a large model capable of generating emissions equivalent to a trans-American flight (Strubell et al., 2019). If sustained, these trends imply substantial increases in energy demand and associated greenhouse gas (GHG) emissions. The increasing use of AI has been associated with broader environmental pressures, ranging from water usage to electronic waste, or e-waste, generated by AI-specific hardware (Falk et al., 2024; Galaz et al., 2021). More recently, investigations have also focused on the indirect or systemic environmental risks of AI, such as rebound effects or path dependencies (Gröger et al., 2025; Luccioni et al., 2025; Uuk et al., 2024), where the entrenchment of AI in socio-technical and ecological systems opens new environmental harms that have until recently not been attributed to AI when assessing the technologies environmental footprint (Kaack et al., 2022).

These tensions warrant further investigation, in particular given AI's rapid diffusion across industrial sectors and society. Early research has investigated the direct environmental impacts of AI, such as energy and water consumption, or e-waste (Li et al., 2025; Patterson et al., 2021; Strubell et al., 2019). More recently, however, investigations have shifted to include the systemic environmental risks of AI (Galaz et al., 2021; Gröger et al., 2025; Luccioni et al., 2025; Uuk et al., 2024). Systemic environmental risks of AI are emergent, cross-sector harms to climate, biodiversity, freshwater, and broader socioecological systems that arise primarily from AI's integration into social, economic, and physical infrastructures, rather than its direct resource use, and that propagate through feedbacks, yielding nonlinear, inequitable, and potentially irreversible impacts. Examples include rebound effects (Luccioni et al., 2025) or path dependencies (Robbins & van Wynsberghe, 2022).

The term "AI" itself is not clearly defined and is often used to refer to different types of technology. For this report, we adopt the definition of Robbins and van Wynsberghe (2022), who refer to AI as "the methodology of creating algorithms driven by the rise of machine learning (ML)," in particular DL. We adopt this as our working definition because many of the methods considered when discussing the environmental impacts of AI are DL-based, and DL-based methods are those that drive the high resource demand of AI (Ligozat et al., 2022; Schwartz et al., 2020; Strubell et al., 2019). In particular, we understand AI as the dominant contemporary paradigm of data-intensive, machine-learning-based systems (Ligozat et al., 2022; Robbins & van Wynsberghe, 2022; Schwartz et al., 2020).

While AI is a subset of the broader Information and Communication Technology (ICT) sector, the development of AI has introduced unique characteristics that differentiate its direct and systemic environmental risks from those of the ICT sector as a whole. First, AI is considered a primary driver of the current expansion in compute infrastructure and the increase in data center electricity consumption (Gröger et al., 2025). The computational requirements of AI have exhibited super-linear growth over the past decade (Bakhtiarifard et al., 2025; Gröger et al., 2025) and show a significantly larger increase than for the ICT sector as a whole (Gröger et al., 2025; Schwartz et al., 2020). Projections indicate that AI is likely to be a key driver of global data-center demand growth. McKinsey (2024) estimates that worldwide demand could grow by 19–22% annually between 2023 and 2030, reaching 171–219 GW in total annual electricity demand by 2030, compared to approximately 60 GW in 2023. In this scenario, AI workloads could account for around 70% of total data center capacity by 2030, with generative AI alone representing roughly 40% of demand. Corporate disclosures report trends consistent with these dynamics. Microsoft reports a 29% increase in total emissions since 2020, largely driven by the expansion of AI infrastructure (Microsoft, 2023). In contrast, Google notes that its overall emissions in 2024 were 51% higher than in 2019, despite efficiency improvements and expanded renewable procurement (Google, 2025). The rapid growth in computing demand has the potential to outpace governance and efficiency gains (Luccioni et al., 2025).

Additionally, AI systems have evolved into general-purpose technology, deployed across various domains (Robbins & van Wynsberghe, 2022; Q. Wang et al., 2024). Their environmental impacts, thus, have the potential to propagate across sectors (Dauvergne, 2022; Robbins & van Wynsberghe, 2022). In fact, AI is often deployed with the intention to alter systems, behavior, and industrial processes, posing a unique systemic risk to the environment (Gröger et al., 2025; Luccioni et al., 2025). These systemic risks are further amplified by the increased concentration within the AI ecosystem, with few frontier models and datasets dominating the landscape (Domínguez Hernández et al., 2024). Finally, the inherent unpredictability of some AI models introduces opacity and new risk vectors due to unforeseen behavior (Robbins & van Wynsberghe, 2022). These risks suggest that while AI is a subset of ICT, its environmental impacts may be disproportionate and more difficult to govern, warranting special investigations into the sustainability of AI (Gröger et al., 2025; Robbins & van Wynsberghe, 2022).

This assessment is based on a narrative review of the existing academic and grey literature, supplemented with expert interviews to inform illustrative case studies. Our contributions are two-fold. First, we offer an overview of the current literature on the systemic environmental risks associated with AI, supplemented by a characterization of the direct impacts. Second, we present an analytical framework for systemic environmental risk to inform the auditing of AI systems with respect to their systemic environmental impacts.

Throughout this report, we focus our investigation on the potential, emerging, and existing environmental risks of AI. Consequently, the report may appear techno-pessimistic at times. However, we acknowledge the potential benefits of AI while offering a risk characterization that enables proactive and precautionary governance and auditing of AI systems, thereby strengthening the understanding and mitigation of potential environmental risks.

We focus the scope of our assessment on the currently dominant technological paradigms: large, deep-learning-based AI systems, such as frontier models, generative AI, and large language models. As such, we acknowledge certain limitations within our approach. The rapid pace of technological development and corporate opacity constrain our ability to provide precise, universally applicable quantifications of environmental harm. Consequently, we recognize that the actual impacts of any AI system are contingent on the context, models, and applications, and thus, our goal is to offer a framework for analysis of individual AI systems.

The analysis in this report is guided by the following research questions:

1. *What are the direct environmental impacts of AI systems across their life cycle?*
2. *What are the key trade-offs between AI model performance and environmental cost, and what are the most prominent strategies for mitigating these costs?*
3. *What are the distinct categories of systemic environmental risks arising from AI's widespread application, and what are their underlying drivers, characteristic pathways, and propagation mechanisms?*
4. *How are the environmental burdens and benefits of the AI life cycle distributed across different regions and populations, and what does this imply for responsibility and accountability?*

Our report is structured as follows. Chapter 2 provides an overview of the direct environmental impacts of AI and their existing mitigation strategies as background and necessary conditions for the systemic environmental impacts. Chapter 3 introduces our proposed taxonomy of systemic environmental impacts of AI and presents three case studies that exemplify both the taxonomy and specific systemic environmental risks associated with AI. In Chapter 4, we consider challenges of distribution and accountability. In Chapter 5, we conclude by offering a synthesis and suggesting priorities for governance, research, and algorithmic auditing.

# 2. Methods

For this study, we use a qualitative mixed-methods approach. We combine a narrative literature review with semi-structured expert interviews. The integration of these methods aims to build an analytical framework based on literature-guided theoretical foundations, supplemented by real-world illustrative examples informed by expert interviews.

**2.1. LITERATURE REVIEW**

We conducted a qualitative, narrative literature review integrating both peer-reviewed research and selected gray literature to inform a concept-building synthesis. We performed literature searches between June and August 2025. While we emphasize recent work to capture current trends in AI, we did not apply any date limits to ensure the inclusion of foundational studies. We followed a tiered search strategy consisting of the following three phases. First, we conducted initial database searches in Scopus and Google Scholar. We employed a Boolean strategy combining keywords across three core conceptual pillars: AI technologies (e.g., "artificial intelligence"), risk pathways (e.g., "systemic risk," "rebound effect"), and environmental domains (e.g., "climate," "sustainability"). The database searches were supplemented by seed paper recommendations from experts active in the field. We manually added literature from prominent authors in the domain by reviewing their publication history. Second, we expanded the initial corpus by using backward and forward citation chasing. We also utilized literature exploration tools, namely Connected Papers, Inciteful, Research Rabbit, and Semantic Scholar, to identify relevant literature that keyword searches might have missed. Third, we conducted targeted literature searches to address identified gaps in specific sub-themes that emerged during the synthesis process.

Sources were screened and selected based on the conceptual relevance to the direct or systemic environmental risks of AI. We included both peer-reviewed and gray literature to balance theoretical rigor with coverage of current developments and real-world applications.
We utilized a narrative synthesis approach using iterative theme development to develop the proposed framework.

We recognize the following limitations to our approach. First, we did not follow a systematic literature review protocol, which may impact both the reproducibility and coverage of our findings. Second, we do not adhere to a strict protocol during the citation chasing phase, which introduces potential inconsistencies in source discovery. Third, the first author performed the selection process, which introduces individual biases in relevance judgments. Fourth, the inclusion of gray literature entails varying degrees of methodological rigor and peer-review processes. Fifth, our focus on English-language literature may exclude relevant work published in other languages, particularly research from non-Western contexts. Finally, our emphasis on recent AI paradigms may have led to the omission of relevant earlier work on computational environmental impacts.

**2.2. EXPERT INTERVIEWS**

To supplement the literature review and to provide illustrative examples, we conducted six semi-structured expert interviews. The interviews were designed to inform the case studies and contextualize our literature-derived framework. The interviews are not intended as primary empirical evidence, but rather serve as an exemplifying or explanatory tool to clarify framework mechanisms described in the review.

We selected case study topics based on prominence in existing literature (oil and gas, agriculture) or based on suggestions that emerged during the initial scoping and outreach (waste management). We identified potential interview partners based on literature review, conference programs, and professional networks according to the following criteria:

- Demonstrated subject-matter expertise (e.g. academic researcher in a relevant subfield)
- Professional experience in a relevant sector or role
- Authorship of relevant peer-reviewed or influential grey literature publications

We interviewed six participants, with each case study having a sample size of n=2. We conducted the interviews online between August and October 2025. Interview durations ranged from 30 to 60 minutes each, following a semi-structured protocol aligned with our proposed framework. This included several more general questions that spanned governance, subject-specific probes, and the perception of AI in their field as a whole. The interviews were consensually recorded, and the recordings were transcribed automatically with manual correction using the open-source transcription software Vibe. Whenever interviews were held in a language other than English, they were translated into English using DeepL. The transcriptions were coded by the first author against deductive categories informed by our framework.

We acknowledge the following methodological limitations. First, the interviews are intended to inform illustrative case studies that do not establish prevalence or generalizable claims. Second, the sample size (n=6) is small, non-representative, and non-random, which may reflect selection and sectoral biases. Third, the reliance on semi-structured interviews means that findings reflect participants' perspectives and experiences rather than objective measurements. The findings should therefore be interpreted as illustrations that complement and contextualize the literature synthesis rather than as empirical evidence.

# 3. The Direct Environmental Footprint of the AI Life Cycle

This chapter addresses the direct environmental footprint of AI. AI is often perceived as an immaterial technology, a narrative supported by the cloud metaphor and frequently criticized for not considering the vast physical infrastructure underlying modern AI systems. We define direct environmental impacts as the immediate consequences of physical processes involved in manufacturing AI hardware and operating AI models, specifically the resource consumption and pollution generated throughout the system's life cycle. These direct impacts create material pressures, such as competition for energy, water, and critical minerals, or the high up-front cost of AI, that form the material basis for the systemic risks analyzed in the following chapter. Thus, understanding these impacts lays the groundwork for a better understanding of the systemic impacts.

The environmental impacts of AI applications have garnered increasing attention, leading to increased investigations into what Schwartz et al. (2020) have termed "Green AI" aimed at mitigating environmental harm. Green AI refers to the research field that seeks to minimize AI's environmental footprint primarily through algorithmic and hardware efficiency. These investigations center on assessing the environmental impacts of AI through life cycle assessments (LCA) to ensure a holistic consideration of these impacts (Dokic et al., 2024; Ligozat et al., 2022; Luccioni et al., 2023; Schneider et al., 2025; Wu et al., 2022). The environmental impacts include electricity and water consumption, GHG emissions, material depletion, e-waste generation, and changes in land use. This chapter synthesizes findings on the primary types and reported magnitudes of these direct impacts, structuring the analysis around three core life cycle stages:

1. **Manufacturing and Infrastructure:** The embodied impacts from raw material extraction and energy-intensive semiconductor fabrication required to build AI hardware.
2. **Operation:** The energy and water consumption incurred by data centers during model training and inference.
3. **End-of-Life:** The environmental impacts associated with the disposal, recycling, and management of obsolete AI hardware.

The analysis focuses on key impact categories consistently highlighted in the literature: energy consumption and associated GHG emissions, water consumption, and the physical footprint of hardware, including material extraction, e-waste, and land use. The chapter concludes with an overview of mitigation strategies and trade-offs, providing context for the accelerating resource demands of the AI industry.

**3.1. MANUFACTURING & INFRASTRUCTURE**

Manufacturing and infrastructure required for AI have considerable environmental costs. These embodied impacts occur along the supply chain of the physical infrastructure powering AI, and include energy, water, and natural resources (Ligozat et al., 2022; Wu et al., 2022). The trend toward more powerful, specialized hardware like Graphics Processing Units (GPUs) and Tensor Processing Units (TPUs) (Morand et al., 2024), and the rapid expansion of AI-driven infrastructure and hyperscale data centers (Gröger et al., 2025; Wu et al., 2022) imply that the embodied footprint is large and increasing. In the following sections, we examine the main components driving the "embodied footprint" of AI.

3.1.1. Raw Materials

AI hardware and data centers require diverse bulk, specialty, and critical materials, each with distinct environmental burdens across extraction, refining, and processing. While there is limited data on the material requirements of AI-specific data centers (Gröger et al., 2025), existing studies estimate the material composition of standard data center rack servers (Leddy et al., 2024). A standard server, weighing approximately 15.8 kg, is typically composed of iron and ferro-alloy metals (12.27 kg) and non-ferrous metals (2.65 kg), with smaller amounts of plastics, precious metals, and other specialized materials (Gröger et al., 2025; Leddy et al., 2024) provides an overview of the primary materials, their use cases in a standard rack server, and the environmental impacts associated with their mining, refining, and production processes.

| MATERIAL/GROUP | ESTIMATED MASS (G)/SERVER | USE | ENVIRONMENTAL IMPACTS |
|---|---|---|---|
| **Aluminum (Bauxite)** | 1,420 (Leddy et al., 2024) | Chassis, Heat Sinks | Deforestation and habitat destruction (Annandale et al., 2021; Giljum et al., 2022), alkaline "red mud" waste (Swain et al., 2022), contamination of water sources (Qi et al., 2022) |
| **Copper** | 1,000 (Leddy et al., 2024) | Wiring, PCBs, Heat Sinks | Acid mine drainage (Johnson & Hallberg, 2005), contamination of soil and water with heavy metals and radioactive materials (Rollog et al., 2019) |
| **Rare Earth Elements** | 0.7-5 (dependent on HDD configuration) | Hard Disk Drive Magnets | Ecosystem destruction, soil and water contamination from toxic and radioactive waste byproducts of refining processes (Schreiber et al., 2016; L. Wang & Liang, 2014) |
| **Gold** | 1.2 (Leddy et al., 2024) | Connectors, PCBs | Deforestation, biodiversity loss, high |

|  |  |  | GHG emissions, mercury and cyanide risks (Esdaile & Chalker, 2018; Trench et al., 2024; Ulrich et al., 2022) |
|---|---|---|---|
| **Tin** | 60 (Leddy et al., 2024) | Solder on PCBs | Seabed and soil disruption, biodiversity loss (Anda et al., 2022; Nurtjahya et al., 2017; Nurtjahya & Agustina, 2015) |
| **Tantalum** | 0.3 (Leddy et al., 2024) | Capacitors | Deforestation, landslides, ecosystem damage, radionuclides (Megevand & Mosnier, 2013; Ojewale, 2022) |
| **Tungsten** | 10 (Leddy et al., 2024) | Vias/interconnects | Tailing acidification, metal leachates (Han et al., 2021; Msumange et al., 2023; Zheng et al., 2024) |

*Table 1: Key Raw Materials in Server Hardware and Associated Environmental Impacts*

The bulk materials, aluminum and copper, have environmental footprints that are critical due to the large volume of these metals represented in servers. Aluminum and its ore, bauxite, are associated with several environmental impacts. One notable environmental impact is habitat and ecosystem disruption. These disruptions are caused by the bauxite mining process, which is often associated with strip mining in tropical regions (Annandale et al., 2021; Giljum et al., 2022). Additionally, the smelting process generates approximately 1.354 tons of highly caustic bauxite residue ("red mud") per 1 ton of alumina globally, on average, which requires long-term storage (Swain et al., 2022). Bauxite residue storage sites are often surface ponds, posing containment risks (Swain et al., 2022). One example of impoundment failure occurred in Ajka, Hungary, in 2010, causing persistent environmental and health concerns (Gelencsér et al., 2011). These risks are also exemplified by case studies demonstrating fluoride plumes migrating from open impoundments, providing evidence of groundwater contamination (Qi et al., 2022). Perfluorocarbons are emitted during smelting, which are potent long-lived greenhouse gases (IPCC, 2019). Similarly, copper mining is associated with the production of acid mine drainage, resulting in environmental pollution (Johnson & Hallberg, 2005). Additionally, copper mining may extract and concentrate radioactive materials (TENORMs) (Rollog et al., 2019), requiring specific management to mitigate environmental damage.

Next to bulk materials, servers rely on what are known as critical and conflict minerals. While they are needed in smaller quantities, they have severe environmental impacts. The first of these are Rare Earth Elements (REEs). Servers typically rely on REEs for magnets in hard-disk drives (Burkhardt et al., 2024). Separation and refinement of REEs rely on extensive solvent-extraction processes and other chemically-intensive steps that have been linked to elevated toxicity, eutrophication, and acidification potentials compared to many metals (Schreiber et al., 2016). A case study found elevated REE levels in atmospheric particulates downwind of tailings in Baotou, China's largest REE industrial base (L. Wang & Liang, 2014).

Another set of minerals within the electronics supply chain of AI are the minerals tin, tantalum, tungsten, and gold, also known as 3TG. 3TG are classified as "conflict minerals" by the EU, meaning that they tend to originate from politically unstable regions and that their mining is considered to finance human rights abuses, as well as support corruption and money laundering (European Commission, 2025). All four of these minerals are linked to environmental harm. Tin mining, for instance, leaves sandy, nutrient-poor soil, therefore causing severe land degradation and potential land use change (Anda et al., 2022). Additionally, studies from the second largest tin producer in the world, the Indonesian province of Bangka Belitung, have shown reduced biodiversity, reduced water quality, seabed changes, an increased mortality index of coral reefs and associated fish, among other environmental harms (Nurtjahya et al., 2017; Nurtjahya & Agustina, 2015).

The Democratic Republic of Congo is a leading producer of tantalum, often mined in the form of coltan (Ojewale, 2022). The mining process is associated with severe environmental damage and frequently occurs without prior environmental impact assessments (Ojewale, 2022). These mining operations are linked to ecosystem and biodiversity disruptions, causing deforestation and forest degradation, as well as land fragility and landslides due to the mining pits (Megevand & Mosnier, 2013; Ojewale, 2022). Moreover, the use of chemicals in mining and processing steps leads to air and water pollution, including the release of radioactive substances into local water bodies (Megevand & Mosnier, 2013; Ojewale, 2022). Indirect and cumulative impacts also arise from associated infrastructure development, such as roads, and an influx of workers (Megevand & Mosnier, 2013).

The environmental risks associated with tungsten are primarily due to contaminants present in the tailings (Han et al., 2021). These contaminants can cause acidic mine drainage (Han et al., 2021; Msumange et al., 2023). The drainage may also contain toxic elements and heavy metals contaminating nearby rivers, groundwater, and soil (Han et al., 2021; Msumange et al., 2023; Zheng et al., 2024).

Gold mining also poses significant environmental concerns. The energy-intensive mining process, as well as the declining ore grade, has led to high GHG emissions, exceeding 100 Mt $CO_2$ annually (Trench et al., 2024; Ulrich et al., 2022). Artisanal and small-scale gold mining is identified as one of the most significant global sources of mercury emissions, releasing over 1,000 tons annually (Esdaile & Chalker, 2018). Finally, gold mining also generates pollution through chemical spillage and acid mine drainage, as well as land degradation resulting from deforestation and extensive surface disturbance (Esdaile & Chalker, 2018; Trench et al., 2024; Ulrich et al., 2022).

Obtaining a precise bill of materials for AI-specific servers remains difficult due to proprietary designs. However, evidence indicates that AI workloads raise chip heat flux and rack power density, driving a transition from air cooling to direct-to-chip and immersion liquid cooling (Alissa et al., 2025; Heydari et al., 2024). These liquid systems incorporate metal-intensive components, including cold plates and heat exchangers that are predominantly made of copper or aluminum (Chinthaparthy et al., 2024; Shahi et al., 2025). AI servers also require more copper for power distribution due to higher rack currents (Chen et al., 2022). Beyond site-level systems, AI accelerators add to the material footprint. Their packaging typically uses nickel-plated copper heat spreaders and vapor-chamber heat sinks. They integrate high-bandwidth memory (HBM) on a silicon interposer with several copper features, including through-silicon vias, copper micro-bumps or copper-to-copper hybrid bonds, and copper redistribution layers (Amin et al., 2009; Huang et al., 2023; Jang et al., 2023; Lee et al., 2025; Zhou et al., 2023). These boards thicken copper planes, connectors, and busbars at the card, backplane, and rack level (IEEE Electronics Packaging Society, 2024). Emerging cradle-to-grave assessments of TPU generations and teardown-based LCAs of GPUs corroborate the material burden of accelerators (Falk et al., 2025; Schneider et al., 2025). Finally, the systems that move air and liquid for cooling often use rare-earth permanent magnets, with additional REEs present where hard disk drive-based storage is deployed (Friebe et al., 2025; Podmiljšak et al., 2024). Taken together, these lines of evidence indicate that AI-specific servers have higher material requirements than standard-use servers, with particularly notable increases in copper, aluminum, and REEs.

### 3.1.2. Semiconductor Manufacturing

A large proportion of natural resources needed by the AI industry are used for semiconductor manufacturing. A process that itself is environmentally harmful, and a large driver of AI's embodied footprint (Hess, 2024; Morand et al., 2024). Semiconductor manufacturing consumes energy and water and emits toxic chemicals and GHGs (Hess, 2024). Front-end wafer processing alone is a complex process, involving more than 50 equipment types, approximately 300 chemicals, and over 1,000 steps (Hess, 2024). Advanced nodes and packaging technologies, e.g., HBM, further add complexity and environmental burdens (Hess, 2024; Schneider et al., 2025). Estimating the environmental impact of semiconductor manufacturing remains difficult, given limited data and high complexity (Hess, 2024). However, a recent first-party cradle-to-grave analysis of TPUs (Schneider et al., 2025) provides the first-of-its-kind analysis of AI accelerators' environmental impact. While this analysis provides a starting point for examining the environmental impacts of AI accelerator manufacturing, it is limited by not including water usage and chemical waste, and by being a first-party report rather than an independent audit.

Semiconductor manufacturing relies on approximately 500 chemicals, many of which have a high Global Warming Potential (GWP) (Hess, 2024). The etching process, for instance, relies on fluorinated gases, known for having a significantly higher GWP than $CO_2$, and accounting for up to 80-90% of fabrication facilities' (fabs') direct emissions (Hess, 2024). Absolute emissions from fluorinated GHGs have remained steady or increased due to growing manufacturing capacities and the increasing complexity of advanced chips (Hess, 2024).

Electricity, often sourced from fossil fuels, is still one of the most significant sources of GHG emissions in chip production (Hess, 2024). The direct energy consumption of AI hardware is challenging to assess due to the limited availability of data. For instance, there is no first-party LCA of GPU manufacturing. However, given the recent TPU LCA, we can consider similar trends for GPUs as well. In particular, as die area increases and node sizes get finer, a trend that enables current accelerator advancements, the GPUs' environmental impact increases (Morand et al., 2024). Schneider et al. (2025) report different TPU generations' emissions from manufacturing and transportation to range from 208 to 585 kg $CO_2$-equivalent ($CO_2$-eq) per TPU chip over a 6-year lifespan. These results are significantly higher than previously used proxy estimates for GPUs (Luccioni et al., 2023; Schneider et al., 2025). They identify higher memory demand for both HBM and Central Processing Unit (CPU) host Dynamic Random Access Memory (DRAM) capacity in newer TPUs as the primary drivers of embodied emissions (Morand et al., 2024; Schneider et al., 2025). Importantly, when considering the entire embodied emissions of the TPU v4i released in 2020 compared to the TPU v6e released in 2024, there is a 1.8-fold increase, nearly doubling the embodied emissions (Schneider et al., 2025). While a similar analysis does not exist for GPUs, the industry trends leading to increased emissions for TPUs are similar for GPUs (Morand et al., 2024; Schneider et al., 2025), indicating potential trends in GPU emissions.

The embodied impacts have been estimated relative to the overall model emissions in several studies. Luccioni et al. (2023) estimate the embodied emissions of the BLOOM model to be 11.2 tons $CO_2$-eq, constituting 22% of total development emissions. Similarly, Mistral AI estimates the hardware embodied emissions of their Large 2 Model to be 2,244 tons $CO_2$-eq, equating to 11% of total emissions[1].

Next to electricity and GHG emissions, semiconductor fabs are also water-intensive, requiring Ultra-Pure Water (UPW) for production and rinsing. Mistral AI estimates its embodied water footprint to be 5% of its total water footprint[2]. Further estimates suggest that a single large fab can use up to 38 million liters of water per day, exceeding many data centers' daily water withdrawals (Hess, 2024). It is estimated that the Taiwan Semiconductor Manufacturing Company (TSMC) withdrew 105 billion liters in 2022 and discharged 71 billion liters of wastewater, containing toxic contamination such as hydrofluoric (HF) acid (Gröger et al., 2025). Gröger et al. (2025) provide two estimates of the global water withdrawals of fabs. The first is extrapolating based on TSMC's market share and water usage, giving an estimate of less than 200 billion liters of water in 2022, likely to have increased in subsequent years due to increased hardware production. The second is based on the estimate of 45 billion liters of water used in fabs in the European Union (EU) in 2024, having a market share of 10%, leading to an estimate of 450 billion liters of water for chip production. These discrepancies might be due to higher water recycling rates in water-scarce regions, such as Taiwan, compared to the EU (Hess, 2024). These numbers consider the impact of the entire ICT sector, including AI. They are thus to be taken as an indication of AI's footprint, given the exponential growth of AI's resource demand (Morand et al., 2024).

---

[1] https://mistral.ai/news/our-contribution-to-a-global-environmental-standard-for-ai

Additionally, semiconductor manufacturing involves and results in chemicals that may cause environmental harm. Hess (2024) provides an overview of these chemicals, such as Per- and Polyfluoroalkyl Substances (PFAS), which are known as "forever chemicals" that do not break down and can accumulate in the environment, with most unused PFAS remaining in wastewater. Furthermore, chemicals used for cleaning, etching, and thin film deposition are energy-intensive in their production, and their disposal poses an environmental risk (Hess, 2024).

In sum, semiconductor manufacturing is a large and fast-growing component of AI's embodied footprint (Morand et al., 2024). More energy-intensive fabrication techniques and higher memory requirements concentrate device-level impacts, which now reach hundreds of kilograms of $CO_2$-eq per accelerator and have nearly doubled across recent TPU generations. These burdens are compounded by significant water withdrawals and toxic wastewater risks, as well as the non-trivial embodiment of the fabs and capital equipment themselves. As electricity for operations potentially decarbonizes, the relative impact of these embodied impacts will increase (Schneider et al., 2025).

### 3.1.3. Data Center Construction

Data center construction is consistently considered to have typically minor environmental impacts, influenced mainly by its impact amortized over its lifespan of approximately 20 years (Schneider et al., 2025). Schneider et al. (2025) report that data center construction accounts for less than 5% of total emissions for the TPU fleet studied. Similarly, Mistral AI reports that data center construction contributes to GHG emissions and water use with less than 1% during their LCA of their Large 2 Model[3]. Further, they report a 1.5% impact on raw material use.

However, other investigations suggest that construction is a carbon-intensive process. Bux et al. (2025), in a LCA of a theoretical data center, use a baseline core-and-shell carbon intensity of 439 kgCO$_2$-eq per square meter. For a typical 8,650 m$^2$ facility, this translates to an upfront emission of nearly 3,800 tCO$_2$-eq.

While emissions, raw material use, and water consumption of constructing data center buildings might not be the largest contributors to AI's environmental footprint, hyperscale data centers, which can span several square kilometers in size (Gröger et al., 2025), constitute a potential form of land-use change. However, design choices, such as choosing a more efficient cooling system, could reduce the required building size by over 16%, thereby lowering the embodied carbon, land use, and material footprint from the outset (Bux et al., 2025). As these data centers are the necessary physical infrastructure for AI, their impact on ecosystems, deforestation, soil sealing, and other environmental concerns must also be considered as an environmental risk associated with AI.

---

[2] https://mistral.ai/news/our-contribution-to-a-global-environmental-standard-for-ai

[3] ibid.

**3.2. OPERATION**

The operation phase of AI systems encompasses the resource requirements during both training and inference. Training is episodic and can be highly energy-intensive, typically involving large clusters of GPUs or TPUs operating for extended periods. Inference, by contrast, requires far less computation per request but occurs at an enormous scale, particularly in the context of generative AI services. As a result, inference is increasingly regarded as the primary contributor to AI's operational environmental footprint (Jegham et al., 2025; Luccioni et al., 2024).

Operational impacts manifest most directly through the electricity consumption of data centers and the associated GHG emissions, as well as through their water usage, both directly for cooling of data centers and indirectly in electricity generation. These impacts are strongly shaped by model size and architecture, hardware efficiency, workload intensity, data center design, and the geographical and infrastructural context in which models are deployed (Adamska et al., 2025; Dauner & Socher, 2025; Li et al., 2023; Zuccon et al., 2023a). While early research into AI's footprint concentrated on the energy requirements of training (Luccioni et al., 2023; Patterson et al., 2021; Strubell et al., 2019), the rapid adoption of generative AI has shifted the center of gravity toward inference. Today, billions of daily queries to generative AI systems such as ChatGPT mean that inference-related impacts accumulate rapidly and are becoming a major sustainability concern.

3.2.1. Energy Consumption

Electricity demand and the resulting GHG emissions are the most widely studied aspect of AI's operational footprint. Early studies highlighted the extreme energy requirements of training Large Language Models (LLMs) (Luccioni et al., 2023; Patterson et al., 2021; Strubell et al., 2019). Strubell et al. (2019), for example, estimated carbon emissions from training common Natural Language Processing (NLP) models, such as BERT and GPT-2, as well as Neural Architecture Search (NAS), by accounting for GPU, CPU, and DRAM power draw alongside Power Usage Effectiveness (PUE) adjustments, excluding inference and infrastructural overhead. They estimated that training BERT on GPU is roughly equivalent to a trans-American flight and fine-tuning a Transformer model with NAS produced approximately 313 $tCO_2$-eq — an amount comparable to the lifetime emissions of five average U.S. cars. Patterson et al. (2021) reported that training GPT-3 consumed around 1,287 MWh of electricity and emitted 552 $tCO_2$-eq, depending on the carbon intensity of the underlying grid. A LCA of BLOOM-176B carried out by Luccioni et al. (2023) further illustrated how boundary choices affect results, with emissions calculated at 24.7 $tCO_2$-eq when only dynamic consumption was considered and at 50.5 $tCO_2$-eq when idle and upstream contributions were included.

As AI systems transitioned into large-scale commercial deployment, inference emerged as the dominant source of operational emissions. Luccioni et al. (2024) conducted a systematic comparison of task-specific and general-purpose models. They found that even relatively small architectures such as BLOOMz-560M required $5.4 \times 10^{-5}$ kWh per inference, while larger models like BLOOMz-7B consumed almost twice as much. The study concluded that multipurpose generative models are orders of magnitude more resource-intensive than narrowly specialized systems when evaluated across diverse tasks. Jegham et al. (2025) benchmarked 30 large language models in commercial data centers and showed that a short GPT-4o query consumed around 0.42 Wh, whereas reasoning-intensive models such as O3 or DeepSeek-R1 exceeded 33 Wh per long prompt. Dauner & Socher (2025) assessed the performance, token use, and $CO_2$-eq across 14 LLMs ranging from 7 to 72 billion parameters.

Their study found large correlation between LLM size, reasoning behavior, token generation and emissions, with larger and reasoning-enabled models achieving higher accuracy but also where responsible for higher emissions.

Recent corporate disclosures further illuminate these dynamics. Google reports that the median Gemini Apps text prompt consumes 0.24 Wh of electricity (Elsworth et al., 2025). Mistral (2025), in collaboration with ADEME (the French ecological transition agency) and Carbone 4, conducted a peer-reviewed LCA of its Mistral Large 2 model. They found that training the models as well as 18 months of its usage generated 20,4 ktCO$_2$-eq. Mistral AI also assessed the impact of the inference for their Large 2 model. They found that a typical 400-token response generates 1.14 gCO$_2$-eq. However, they did not publish the energy consumption. While these values may appear small in isolation, their significance becomes clear when multiplied across hundreds of millions or billions of daily queries. OpenAI reports that ChatGPT alone processes around 2.5 billion prompts per day (Chatterji et al., 2025).

3.2.2. Water Consumption

Water demand has only recently gained attention as a key element of AI's operational footprint. Data centers consume water both directly, mainly for cooling, and indirectly, via electricity generation (Hoffmann et al., 2025; Ristic et al., 2015a). Cooling alone can represent 30-40% of a facility's total energy use, underscoring the close coupling between energy and water footprints (Q. Zhang et al., 2021). Case studies already suggest substantial demands during model training. Li et al. (2023) estimated that training GPT-3 in Microsoft's U.S. data centers could consume a total of 5.4 million liters of water, including a total of 700,000 liters of on-site water consumption for cooling. The same study projected that global AI-related withdrawals could reach 4.2-6.6 billion cubic meters annually by 2027—equivalent to 4-6 times the yearly consumption of Denmark or half that of the United Kingdom.

Inference, too, carries significant water implications. Shumba et al. (2024) estimated that generating a ten-page report with LLaMA-3-70B consumed around 0.7 liters of water, whereas using GPT-4 for the same task could require as much as 60 liters, depending on the infrastructure and energy mix. Zuccon et al. (2023) demonstrated that larger information retrieval models disproportionately increase water use, while local factors such as seasonality, cooling technology, and grid composition exert strong additional influence.

Industry disclosures provide complementary insights. Google reports that the median Gemini Apps text prompt requires about 0.26 mL of on-site water, equivalent to five drops, although this calculation only includes direct water consumption for cooling and excludes off-site use in electricity generation (Elsworth et al., 2025). In contrast, Mistral (2025) reports that a 400-token query to its "Le Chat" assistant consumes 45 mL of water, including both direct and indirect consumption.

Corporate sustainability reports confirm that overall water use is rising in step with AI adoption. Microsoft documented a 34% increase in water withdrawals between 2021 and 2022, reaching approximately 6.4 million cubic meters, while Google reported a 20% increase during the same period (Google, 2023; Microsoft, 2022a). Across the literature, results vary by orders of magnitude due to differences in scope and methodology, yet the emerging consensus is clear: water consumption has become a critical environmental risk factor during both training and inference (Hoffmann et al., 2025).

**3.3. END-OF-LIFE**
The End-of-Life (EOL) phase of AI systems is another type of embodied impact. It covers dismantling, refurbishment, material recovery, and disposal and includes retirement reverse logistics (Schneider et al., 2025). In particular, the disposal and material recovery can contribute to the environmental impact of AI (Morand et al., 2024). While operational emissions typically dominate, they can be offset. In contrast, the materiality of the EOL phase is less easily mitigated and, thus, can rise relative to operational cost if the energy supply is decarbonized (Schneider et al., 2025; Wu et al., 2022).

The growth in AI is projected to significantly contribute to global waste streams, with estimates suggesting between 1.2 and 5 million tons of additional e-waste could be generated by generative AI alone from 2020 to 2030 (P. Wang et al., 2024). Gröger et al. (2025) estimate that data center expansion between 2023 and 2030 may add 4.2 million tons of e-waste to the existing global annual e-waste of 62 million tons. Additionally, the waste stream from semiconductor manufacturing reportedly doubled between 2016 and 2024 (Hess, 2024). It includes hazardous chemical waste, wastewater, and substantial packaging waste, and, therefore, can contribute to ecosystem damage if mismanaged (Hess, 2024).

Recycling and refurbishment efforts are also lacking. While one EU waste disposal stakeholder reported that of the servers collected, 81% of the mass is reused or recycled (Leddy et al., 2024), this is likely to prioritize bulk materials over critical raw materials, where manual recovery is more difficult and therefore often not commercially viable (Gröger et al., 2025; Leddy et al., 2024). Moreover, servers are frequently retired before their technical EOL, and data security concerns still drive some operators towards physical destruction rather than refurbishment efforts (Leddy et al., 2024). Finally, estimates suggest that only about one-fifth of global e-waste is formally collected and documented, with a collection and recycling rate of 22.3% in 2022, projected to decline further to 20% by 2030 (Gröger et al., 2025).

**3.4. TRADE-OFFS & MITIGATION STRATEGIES**
AI's environmental footprint involves complex trade-offs across its life cycle stages, requiring a balance between performance gains and resource costs. Improving on state-of-the-art accuracy often incurs exponentially higher computational expense, where minor accuracy gains are achieved at a disproportionately high energy and material cost. For example, achieving linear improvements in model performance can require exponentially larger models or datasets (Schwartz et al., 2020). Beyond a certain performance threshold, further improvement may not be justified by the environmental cost (Schwartz et al., 2020). Thus, the principle of "good-enough" models implies considering model performance in the context of its environmental impacts, rather than aiming for state-of-the-art performance at any cost. In the following, we examine key trade-offs and mitigation approaches at each stage of the AI life cycle, highlighting strategies for "Green AI" that balance innovation with sustainability.

In the literature on sustainable computing, mitigation strategies are broadly categorized along two complementary axes: efficiency and sufficiency (Görücü et al., 2025; Gröger et al., 2025; Heilinger et al., 2024; Weber et al., 2023; Wright et al., 2025). As previously discussed in the literature, the focus on the dominant paradigm of efficiency, such as performance per watt, is insufficient (Görücü et al., 2025; Wright et al., 2025). Instead, the integration of sufficiency as a coequal principle is needed.

Sufficiency extends optimization by evaluating the scale and necessity of the output itself. Next to questioning model efficiency, sufficiency implies considering the minimal viable solution with the minimal environmental cost for the task. In this chapter, we proceed by considering current trade-off approaches for AI's environmental impacts, aiming to cover both efficiency and sufficiency considerations. An overview of trade-offs and mitigation strategies by life cycle stage is provided in Table 2.

| Life Cycle Stage | Primary Environmental Impacts | Key Trade-offs | Mitigation Strategies (Efficiency-Based) | Mitigation Strategies (Sufficiency-Based) |
|---|---|---|---|---|
| **1. Hardware Manufacturing** | Embodied carbon & energy; resource depletion; water consumption; toxic waste | Operational efficiency vs. manufacturing complexity & embodied carbon | More efficient chip architectures; sustainable semiconductor manufacturing processes | Circular design; Right to Repair legislation; extended hardware life cycles |
| **2. Data** | Energy for data storage, transfer, and processing; hardware footprint of storage infrastructure | Dataset size and quality vs. model accuracy and generalizability | More efficient storage hardware and data compression algorithms | Dataset distillation & coreset selection; active learning; data life cycle management |
| **3. Model Training** | Electricity consumption; water consumption; carbon emissions | Model accuracy and capability vs. computational cost | Model compression; efficient architectures; transfer learning; efficient compute infrastructure | "Good enough" model selection; early stopping; data-centric approaches to reduce training needs |
| **4. Model Inference** | Cumulative energy consumption; network energy for data transmission | Inference latency and accuracy vs. energy per query; cloud efficiency vs. edge device proliferation | Optimized models for inference; efficient hardware; efficient compute infrastructure | Adaptive computation; selecting the smallest model sufficient for the task; query batching |
| **5. Hardware EOL** | E-waste; leaching of toxic materials; loss of valuable resources | Rapid innovation and performance gains vs. hardware longevity and product obsolescence | Improved recycling efficiency; advanced material recovery processes | Hardware reuse and refurbishment; secondary markets for used equipment; Right to Repair |

*Table 2: AI Life Cycle Stages: Environmental Impacts, Trade-offs, and Mitigation Strategies*

### 3.4.1. Infrastructure & Manufacturing

The trade-off during the infrastructure and manufacturing phase of AI systems lies between operational efficiency and the carbon cost of manufacturing (Różycki et al., 2025). Advanced AI accelerators can significantly improve measures such as performance per watt, yet they require more complex and energy-intensive manufacturing (Różycki et al., 2025; Schneider et al., 2025). These trade-offs become increasingly important as the operational environmental impacts of AI systems reduce (Schneider et al., 2025).

To address the environmental footprint of AI infrastructure, one efficiency-based mitigation strategy is to develop more energy-efficient hardware (Bolón-Canedo et al., 2024; Goel et al., 2024; Różycki et al., 2025). For example, specialized accelerators, such as TPUs, have been shown to offer better performance per watt than traditional GPUs (Elsworth et al., 2025; Różycki et al., 2025). Moreover, as Schneider et al. (2025) show, newer architectures of the same accelerator have improved compute carbon intensity, a measure they introduce considering $CO_2$-eq per Floating-Point Operation (FLOP). Thus, more efficient hardware performs the same computations with lower energy use, thereby lowering the operational carbon emissions of AI. Another efficiency-based mitigation strategy is improving the impact of chip fabrication (Yeboah et al., 2025). These efforts include investing in cleaner energy, as well as initiatives for water and material recycling (TSMC, 2024; Yeboah et al., 2025). The largest chip manufacturer, TSMC, reports intensified efforts in water reclamation. They introduced reclaimed water into 5- and 3-nanometer processes at the Taiwan plant in 2024, reporting a 31% reduction in tap water consumption (TSMC, 2024). Additional efforts focus on reducing toxic byproducts of chip manufacturing. Again, TSMC (2024) report optimizations to ammonia-nitrogen wastewater treatment, resulting in a 30% decrease in chemical agent consumption while reducing waste generation and conserving energy.

Efficiency, however, does not address hardware refresh cycles and upgrade pressures. Evidence suggests that typical server refresh times may be slowing. However, demand and platform-level drivers can still push frequent upgrades (Weber et al., 2023; Wright et al., 2025). Sufficiency measures, on the other hand, call for prolonged hardware life cycles and reducing the need for new devices (Goel et al., 2024; Gröger et al., 2025). One strategy to achieve sufficiency is by designing for longevity and repair. By ensuring hardware is modular and repairable, the production and e-waste cycle can be slowed, thereby increasing the likelihood of longer hardware lifespans (Goel et al., 2024). Similarly, circular design principles aim to design hardware with EOL considerations in mind (Goel et al., 2024; Yeboah et al., 2025). For instance, eco-design that enables easy disassembly, repair, and modular upgrades can improve recyclability and extend lifetimes (Goel et al., 2024). Producer-responsibility and buy-back programs can further increase take-back and recycling (Goel et al., 2024). Finally, promoting secondary markets for used equipment and refurbishment helps ensure that existing hardware is continually utilized rather than entering the waste stream (Goel et al., 2024; Gröger et al., 2025). There is growing industry and policy support for refurbishing data center equipment and consumer devices for second-hand sale, which both reduces e-waste and makes computing more accessible (Goel et al., 2024; Gröger et al., 2025). By reusing hardware and avoiding needless upgrades, sufficiency-centered approaches lessen the demand for constant manufacturing, thereby reducing the embodied carbon and resource depletion associated with infrastructure.

### 3.4.2. Operation

During operation, each sub-stage, such as data, training, and inference, is subject to unique trade-offs between performance and resource use.

***Data:*** Approaching efficiency from a data perspective has garnered increasing attention in recent years, giving rise to the concept of data-centric Green AI (Salehi & Schmeink, 2024; Yu et al., 2024). Improving hardware and software for data storage can reduce the operational impact of AI (Gröger et al., 2025). For instance, using energy-efficient storage media, such as modern Solid State Drives (SSDs), can reduce electricity usage (Goel et al., 2024). Additionally, data storage tiering, where frequently accessed data is stored in energy-intensive, high-performance storage, and less often used data is migrated to slower, less energy-intensive storage media, can help optimize energy usage (You et al., 2020). Furthermore, employing data compression and deduplication can reduce data size, thereby lowering the operational cost required to store data (Prajapati & Shah, 2022).

Approaching data sufficiency, on the other hand, considers how much data needs to be collected and retained. Techniques such as dataset distillation or coreset selection aim to curate small, information-rich subsets of data that achieve nearly optimal model performance compared to a large, full dataset (Yu et al., 2024). Moreover, active learning can be used to inform data selection, considering only labeling and storing those data points that are most informative (Salehi & Schmeink, 2024). This avoids collecting large datasets by focusing on informative, high-quality samples (Salehi & Schmeink, 2024). Both approaches yield a smaller dataset, thereby improving storage costs and training time while maintaining the AI system's performance (Salehi & Schmeink, 2024; Yu et al., 2024). Finally, considering data life cycle management, such as regularly deleting or archiving data, can also reduce the environmental footprint of AI (Różycki et al., 2025). All measures aim to maintain smaller data repositories, focusing on quality over quantity and lowering the continuous cost of data storage.

***Training:*** Training efficiency is the most thoroughly investigated mitigation strategy of AI's environmental impact. The primary trade-off considered is between model performance, such as accuracy, and environmental impact, including energy use (Schwartz et al., 2020; Verdecchia et al., 2023). Efficiency-based mitigation strategies typically aim to reduce computational cost while retaining model performance (Bolón-Canedo et al., 2024; Różycki et al., 2025; Schwartz et al., 2020). One widely followed approach is that of algorithmic efficiency (Bolón-Canedo et al., 2024; Różycki et al., 2025). Low-precision training is one example of such a mitigation strategy. AI models are trained with lower numeric precision, commonly using 16-bit or even 8-bit floating-point precision over 32-bit, which can reduce the compute and power required for each operation (Nenno, 2024; Różycki et al., 2025). Another avenue of efficient training is selecting an efficient model architecture (Bakhtiarifard et al., 2024; Bolón-Canedo et al., 2024; Różycki et al., 2025). Research has led to model architectures, such as EfficientNet or MobileNets, that are designed to achieve a given performance with fewer parameters and thus lower energy requirements (Bakhtiarifard et al., 2024). Similarly, NAS can be achieved with energy-aware strategies that both reduce the environmental impact of the search itself and the resulting AI models (Bakhtiarifard et al., 2024; Y. Zhao et al., 2024). For retraining or fine-tuning, the environmental cost can be reduced by employing model compression methods, such as pruning or model distillation, which achieve similar performance with reduced model complexity (Bolón-Canedo et al., 2024; Różycki et al., 2025). Transfer learning is another mitigation strategy, where large pre-trained models are fine-tuned on smaller task-specific datasets (Liu & Yin, 2024; Różycki et al., 2025). Transfer learning relies on one large pre-trained model, rather than fully training large models for each new task, reusing prior computational cost to lower the new training cost, a strategy widely adopted in NLP and Computer Vision (Liu & Yin, 2024; Różycki et al., 2025; Schwartz et al., 2020). Finally, considering the system-side efficiency is a strategy to mitigate training cost. Using compute infrastructure with low carbon intensity or scheduling training to run in regions or times where carbon intensity is low can reduce the environmental footprint of AI training (Różycki et al., 2025; Wright et al., 2025; Y. Zhao et al., 2024).

There are several sufficiency-based mitigation strategies as well. The main sufficiency approach is considering the necessity of large AI models and long training regimes for a given task (Bolón-Canedo et al., 2024; Gröger et al., 2025). That is to say, considering smaller models and shorter training resulting in "good enough" approaches that are significantly cheaper to train, while potentially slightly trailing the state-of-the-art performance (Bolón-Canedo et al., 2024; Gröger et al., 2025). Additionally, a specific resource budget can be set in advance for training. This restricts unconstrained experimentation that might utilize exponential resources for diminishing returns and establishes a justified resource allocation for performance gains. Early stopping can be considered a form of both, where training is stopped after the model's performance on a validation dataset does not improve significantly further (Różycki et al., 2025). Finally, considering data-centric approaches can also be a sufficiency measure for AI model training (Salehi & Schmeink, 2024; Verdecchia et al., 2023). Rather than using large models with large datasets, one can focus on constructing a high-quality dataset that is potentially smaller, allowing for the training of a smaller model (Nenno, 2024; Salehi & Schmeink, 2024; Yu et al., 2024). Dataset distillation or coreset selection can also be employed to reduce training costs (Yu et al., 2024).

*Inference:* Both trade-off and efficiency-focused mitigation strategies for inference are similar to those for training. For instance, quantization is also commonly used to make inference more efficient (Barbierato & Gatti, 2024; Bolón-Canedo et al., 2024). By quantizing the weights of AI models after training, the inference cost and memory usage can be significantly reduced, allowing for the use of specialized low-precision hardware (Bolón-Canedo et al., 2024; Różycki et al., 2025; Wu et al., 2022). AI models can be further optimized for efficient inference by utilizing techniques such as model distillation or model pruning post-training (Barbierato & Gatti, 2024; Elsworth et al., 2025; Różycki et al., 2025). In addition, specialized, efficient hardware, such as Application-Specific Integrated Circuits (ASICs) or Field Programmable Gate Arrays (FPGAs) optimized for a specific model's operations (Barbierato & Gatti, 2024; Wu et al., 2022), or leveraging approaches that throttle power when not fully utilized, can enhance inference efficiency (Elsworth et al., 2025). Query batching, on the other hand, offers efficiency gains by ensuring the accelerator infrastructure is always optimally utilized (Elsworth et al., 2025). Finally, considering efficient compute infrastructure is a crucial factor in reducing the environmental impact of inference (Elsworth et al., 2025; Różycki et al., 2025; Wright et al., 2025).

Several mitigation strategies can be employed, with a focus on sufficiency. One approach is to adjust the computational effort based on the input dynamically (Elsworth et al., 2025). In LLMs, for instance, techniques such as Mixture-of-Experts (MoE) or Speculative Decoding enable a model to generate a response for a simple query using only a fraction of its layers (Elsworth et al., 2025). MoE, for example, activates a small subset of a large model required explicitly for a prompt, reducing computations and data transfer by a factor of 10-100x (Elsworth et al., 2025). Another approach is to use a group of models, ranging from efficient to computationally expensive, and aim to route the input to the smallest model that is adequate for the task, as seen in OpenAI's recent introduction of GPT-5[4]. Finally, query batching and load management can enhance sufficiency, where queries are batched to optimize infrastructure utilization (Elsworth et al., 2025). Considering this further, job scheduling, i.e., waiting to execute batched queries until the carbon intensity of a given data center is more optimal, could be employed as a sufficiency measure to reduce the environmental footprint of inference (Wright et al., 2025; Wu et al., 2022).

### 3.4.3. End-of-Life

Trade-offs in the EOL phase of AI mainly consider hardware longevity and innovation in the form of new hardware.

Efficiency-focused mitigation strategies are primarily two-fold. First, improving upon the recycling and disposal processes can reduce the environmental harm of hardware at its EOL (Zhuk, 2023). Second, improving material recovery rates could decrease the need for virgin materials (Goel et al., 2024; Yeboah et al., 2025). Moreover, improved recovery rates may make recycling more economically viable, thereby incentivizing recycling over waste exports (Goel et al., 2024; Leddy et al., 2024). Standardizing specific components or utilizing more recyclable materials can also lead to a more efficient recycling process (Leddy et al., 2024; Zhuk, 2023).

While efficient recycling is essential, sufficiency considers prolonging the hardware's life cycle to increase amortized cost before it is discarded (Gröger et al., 2025; Wu et al., 2022). Sufficiency at EOL centers on extending the usable life of hardware and diverting devices away from disposal (Gröger et al., 2025; Zhuk, 2023). One primary strategy is the reuse and refurbishment of hardware (Zhuk, 2023). Older hardware, such as last-generation servers, could be used as secondary devices for less critical workloads (Kaack et al., 2022; Leddy et al., 2024). They could be refurbished and resold to extend the hardware's lifespan (Leddy et al., 2024; Zhuk, 2023). This would also benefit access to AI hardware, as low-cost, refurbished hardware would become more widely available, allowing for more equitable access to compute infrastructure (Bakhtiarifard et al., 2025; Leddy et al., 2024).

---

[4] https://academy.openai.com/public/clubs/work-users-ynjqu/resources/intro-gpt-5#

By delaying the entry of these materials into the waste stream and reducing the demand for new units, environmental harm can be minimized (Goel et al., 2024; Zhuk, 2023). Relatedly, fostering secondary markets for used components enables upgrades to occur circularly (Leddy et al., 2024; Zhuk, 2023). Increasing the facilitation of this exchange could imply fewer total devices need to be produced (Goel et al., 2024; Zhuk, 2023). The Right to Repair and Right to Reuse initiatives further support the idea that consumers and third-party technicians should be able to repair, resell, refurbish, and reuse hardware without facing legal or technical barriers (Leddy et al., 2024). For instance, laws that require manufacturers to provide spare parts and repair information make it feasible to fix older devices and keep them in service longer. Additionally, circular economy programs encourage manufacturers to take back old products and either remanufacture them or extract components (Leddy et al., 2024; Zhuk, 2023).

**3.5. CONCLUSION**

In this chapter, we examined the materiality of AI's environmental footprint, from mineral extraction to EOL waste management. It highlights two key environmental challenges: the initial footprint resulting from manufacturing complex hardware and the ongoing energy and water usage associated with AI models deployed at scale. A key takeaway is the interplay between these factors. As operations become more efficient and energy sources become less carbon-intensive, the relative impact of embodied carbon and material depletion grows. This highlights the inadequacy of mitigation strategies that focus solely on operational efficiency, underscoring the necessity of implementing sufficient measures to address hardware life cycles and reduce overall consumption.

A core limitation of this chapter lies in the quantification of these impacts, which is consistently hindered by corporate opacity and rapid technological advancements that render data obsolete. This uncertainty is itself a critical finding, complicating governance efforts. These direct impacts are not isolated issues. These form the material and economic basis for the systemic environmental risks discussed in the next section.

# 4. Systemic Environmental Risks

In this chapter, we examine the systemic environmental risks of AI. While the direct impacts discussed in the previous chapter have received increasing attention, systemic risks remain comparatively underexplored. Systemic environmental risks of AI are emergent, cross-sector harms to climate, biodiversity, freshwater, and broader socioecological systems that arise primarily from AI's integration into social, economic, and physical infrastructures, rather than its direct resource use, and that propagate through feedbacks, yielding nonlinear, inequitable, and potentially irreversible impacts.

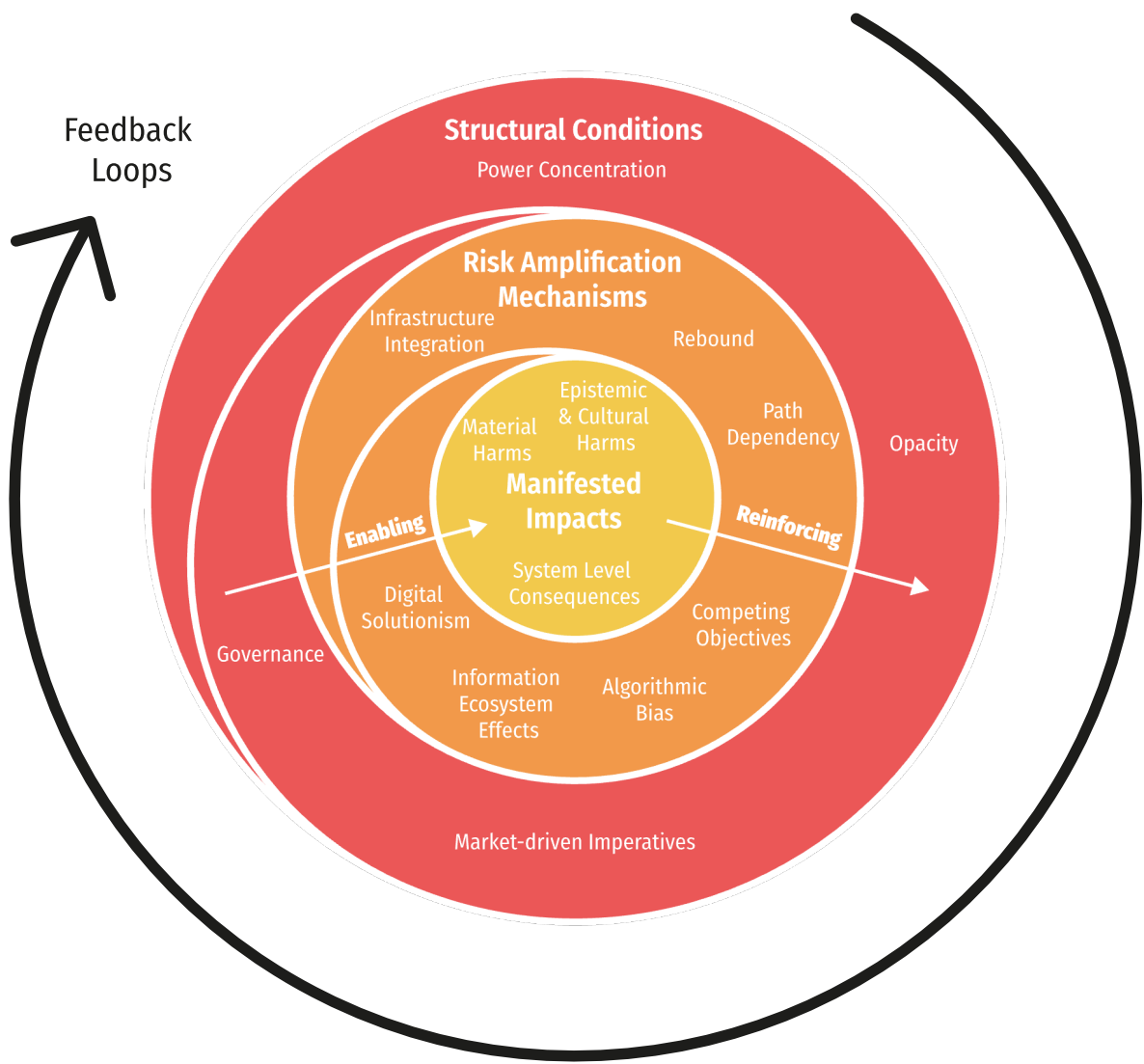

*Figure 1: Illustration of the analytical framework for the systemic environmental risks of AI.*

Although many technologies entail systemic environmental risks, we argue that AI is particularly well-positioned to create and amplify them due to a combination of the following:

- Pervasiveness: AI's applicability extends beyond a single sector, giving it the potential to reshape resource-intensive industries and practices at scale.
- Pace and Scale of Deployment: The capabilities of frontier models enable rapid, large-scale deployment that can outpace the development of effective environmental governance and propagate harms along global value chains.
- Technical Opacity: The internal logic of many modern ML systems is inherently difficult to audit or predict, a challenge compounded by corporate secrecy that can conceal embedded environmental externalities.
- Propagation and Correlation Risks: The current paradigm of developing applications based on a few foundation models, large-scale datasets, or technical design choices creates new vectors of risk propagation, allowing environmental harms to cascade across previously disconnected socio-ecological systems.

Given the nascent nature of the field, there is currently no established definition of systemic environmental risks. Based on definitions given in current literature (Falk et al., 2024; Galaz et al., 2021; Gröger et al., 2025; Luccioni, Strubell, et al., 2025; Uuk et al., 2024), we adopt the following, more comprehensive, working definition: Systemic environmental risks of AI are large-scale, cross-sector harms to climate, biodiversity, freshwater, and other Earth system processes that arise less from AI's direct resource footprint than from emergent, system-level interactions created when the AI life cycle is embedded in sociotechnical and ecological systems. Under competitive market pressures and weak or misaligned governance, these risks propagate through indirect channels and feedback loops, producing nonlinear, hard-to-attribute, potentially irreversible, and inequitable impacts that can accumulate and cascade across sectors and geographies.

To operationalize this definition, we propose a three-level framework that acknowledges the interconnectedness and reinforcing feedback loops between its components. This structure simplifies analysis while emphasizing the need for a holistic approach. The three levels are:

- **Structural Conditions:** The economic, political, and technical contexts shaping AI
- **Risk Amplification Mechanisms:** The specific pathways through which AI propagates environmental harm.
- **Manifested Impacts:** The observable ecological and social consequences.

The following sections employ this framework to analyze the sources, pathways, and impacts of AI's systemic environmental risks. To ground this analysis, we integrate illustrative quotes from expert interviews throughout the chapter, supplemented by case studies based on the expert interviews.

### 4.1. STRUCTURAL CONDITIONS
In this section, we introduce the structural conditions of systemic environmental risks of AI. While these structural conditions enable systemic risks on their own, additional risk pathways emerge from their interconnectedness and feedback loops. We highlight potential feedback loops at the end of each section.

#### 4.1.1. Power Concentration & Asymmetries
In this section, we explore how the concentration of technological and economic power in a few actors and regions creates structural imbalances. The core issue is a disconnect between those who benefit from AI and those who bear its environmental costs. This disconnect creates a condition where harmful practices can be perpetuated by distancing decision-makers from their consequences.

***Unequal Distribution of Burdens and Benefits:*** Systemic environmental risks of AI are fundamentally enabled by the unequal distribution of burdens and benefits (Domínguez Hernández et al., 2024; Falk et al., 2024; Galaz et al., 2021). This division predominantly manifests along the lines of social groups and geographical regions, with those mainly benefiting being located in High-Income Countries (HICs), and those disproportionately affected by the environmental risks being situated in Low- to Middle-Income Countries (LMICs) (Bakhtiarifard et al., 2025; Domínguez Hernández et al., 2024; Sander, 2025). This disparity results in what we term a "cognitive buffer", distancing the developers and those benefitting from AI systems from its environmental consequences (Dauvergne, 2022; Robbins & van Wynsberghe, 2022). The displacement of environmental costs creates "sacrifice zones" that bear the cost of AI innovation while receiving minimal benefits, and often play a critical role in global environmental stability (Galaz et al., 2021; Lehuedé, 2025; Murdock, 2025). The cognitive buffer enables this exploitation by making environmental consequences psychologically and physically distant from those who create them and those with the power to address them. The cognitive buffer insulates decision-makers from the negative consequences of their actions, making environmentally harmful practices and systemic environmental risks more likely to occur and persist.

Compounding this is a lack of transparency and disclosure from AI developers and corporations, obfuscating the full scope of environmental impacts (Luccioni, Strubell, et al., 2025; Sampatsing et al., 2025; UNEP, 2024), while emphasizing AI's positive contributions (Dauvergne, 2022; Heilinger et al., 2024). The lack of transparency, combined with the geographical and social disconnect, fosters what Lucivero (2023) calls a "responsibility gap", hindering the ability of legal frameworks and policy initiatives to effectively address the systemic environmental risks of AI (Falk et al., 2024; Haider et al., 2022).

***Scale Requirements and Power Concentration:*** The need for compute, data, and infrastructure at scale erects barriers to access across modern AI development (Bakhtiarifard et al., 2025). These structural demands have led to a consolidation of power within a limited set of firms and institutions, primarily based in HICs, who possess the financial and technical means to operate at scale (Bakhtiarifard et al., 2025; Domínguez Hernández et al., 2024). The resulting market concentration affords these actors disproportionate influence over the framing and shaping of ethical, policy, cultural, and educational frameworks that govern AI (Bakhtiarifard et al., 2025; Brevini, 2021; Schütze, 2024a). An illustrative example of power concentration can be drawn from our case studies. One expert commented on the power concentration of the agricultural-technology sector that "I can see a world in which it is not a global public good because the private sector has access to it in a way that the public sector will never be able to reach", exemplifying how power is concentrated in corporations. This concentration facilitates the global diffusion of AI systems that encode the values, assumptions, and epistemic frameworks of their developers, while marginalizing alternative knowledge systems and local expertise (Lehuedé, 2025; Nost & Colven, 2022). These systems are often presented as neutral or universally applicable, despite being trained on context-specific data that may be misaligned with local environmental or social realities (Nost & Colven, 2022). This imposition undermines local sustainability initiatives and reinforces dependencies on external solutions that may be environmentally inappropriate for the specific context (Schütze, 2024b).

The concentration of power also enables dominant actors to shape AI's narrative framing, thereby influencing policy and public discourse (Brevini, 2021). AI is promoted as an essential and inevitable solution to global challenges, such as climate change, often emphasizing technical efficiency and data-driven optimization, while frequently downplaying the environmental externalities of AI systems themselves (Cellard et al., 2025; Sander, 2025).

***Feedback to Governance Failures:*** Power concentration and unequal distribution of benefits and burdens directly undermine governance by ensuring that those with the most significant influence over AI development have strong incentives to resist environmental regulation. Concentrated actors use their resources to lobby against restrictive policies, fund favorable research, and capture regulatory processes (Ebert et al., 2025; Hendrycks et al., 2023; Sander, 2025), creating the governance failures examined in the next section. The development of the EU AI Act provides a concrete example of this feedback loop. During the final "trilogue" negotiations, stronger environmental legislation proposed by the European Parliament was diluted, reflecting the influence concentrated actors can have on regulatory processes (Ebert et al., 2025; Hacker, 2024).

### 4.1.2. Governance

Current AI governance exhibits flaws that create a permissive environment for systemic environmental risks. Rather than proactively addressing the potential risks posed by AI, governance is characterized by a reactive approach and a reliance on ineffective voluntary measures, and is often undermined by geopolitical competition.

***Regulatory Lag:*** AI technology is developing at an exponential rate, with estimates suggesting that the computational resources needed to create prominent AI technology have been doubling every 3.4 months for more than a decade (Bakhtiarifard et al., 2025). While there are some proposals for AI regulation, such as the EU AI Act, regulatory legislation lags behind the rapid progress of AI (Ojanen, 2025). The reactive nature of regulation, in combination with the widespread adoption and rapid development, means that AI's dependencies and environmental consequences can become deeply embedded and difficult to reverse (Ojanen, 2025; Robbins & van Wynsberghe, 2022). This allows systemic environmental risks to develop before appropriate regulatory measures can be taken (Robbins & van Wynsberghe, 2022).

Current regulatory efforts are criticized for prioritizing human safety and individual risk over environmental harm (Falk et al., 2024; Gröger et al., 2025). Environmental sustainability is typically only considered tangentially or as a voluntary consideration for AI developers (Galaz et al., 2021; Hacker, 2024; Lucivero, 2023). By prioritizing immediate human utility, this anthropocentric focus systematically risks failing to adequately address systemic environmental harms (Falk et al., 2024; Hacker, 2024; Thomas et al., 2025).

***Self-Regulation:*** The regulatory deficiencies lead to governance gaps, which strengthen the reliance on industry self-regulation, a dynamic often reinforced by private-sector influence and risks of regulatory capture from large technology companies (Lucivero, 2023; J. Zhao & Gómez Fariñas, 2023). While governance measures are increasing, those that exist mostly remain as guidelines and recommendations, lacking clear implementation and enforcement mechanisms (Galaz et al., 2021; Lucivero, 2023). One expert on waste-robotics describes the regulatory landscape as "For the AI part, nothing. […] when it comes to if our AI was such that the sorting accuracy isn't really matching the environmental permits or something like that, we don't have any, there is no standardization there." Historically, voluntary commitments to address climate change have often failed to yield meaningful outcomes, suggesting that this approach may be inadequate for AI governance (Hacker, 2024).

Current "Green AI" initiatives exemplify this pattern. These initiatives lack independent verification and transparency, have a narrow focus on carbon emissions, and are frequently secondary to profit-driven incentives (Cellard et al., 2025; Luccioni, Strubell, et al., 2025; Sampatsing et al., 2025). The critique of Corporate Social Responsibility (CSR) programs often suggests that climate goals are only pursued if they align with existing business incentives (Dauvergne, 2022). That is to say, "Green AI" initiatives may be driven more by cost-saving than genuine environmental concern, allowing resource-intensive practices to continue (Dauvergne, 2022; Nost & Colven, 2022). This has led to the strategic response of creating the appearance of environmental responsibility to gain economic and political benefits by being perceived as sustainable, thereby obscuring AI's true environmental costs (Dauvergne, 2022; Heilinger et al., 2024; Nost & Colven, 2022). One expert described this corporate reputation management in practice, noting how "Microsoft renamed their page from 'AI in Oil & Gas' to 'AI in Energy.'" after facing backlash.

The self-regulatory model is further weakened by focusing on carbon emissions and neglecting the full environmental impact of AI, which enables the externalization and obfuscation of environmental costs, particularly systemic environmental harms (Cellard et al., 2025; Dauvergne, 2022; Luccioni, Strubell, et al., 2025; Wright et al., 2025). This is compounded by what Falk et al. (2024) call the "attribution problem": while aggregate impacts may be possible to quantify, the complex and interconnected nature of these systems makes it challenging to attribute them to a single model, company, or application. This complexity affords corporations plausible deniability and renders targeted regulation challenging.

**Geopolitical Competition:** A further complication in the development of regulation to ensure that AI systems are designed sustainably is a newly invigorated geopolitical competition for "AI supremacy". Francisco (2023) liken this competition to the "space race" where "winners will take it all" and, as a result, AI development is widely viewed as a central contest for global influence and power (Brevini, 2020; Francisco, 2023). This is also echoed recent in statements and regulatory efforts, with Chinese leader Xi Jinping being quoted describing AI as "a strategic technology leading a new round of scientific and technological revolution and industrial transformation"[5] and the Trump Administration stating that "it is the policy of the United States to sustain and enhance America's global AI dominance to promote human flourishing, economic competitiveness, and national security."[6]. The focus on national security is further exemplified by the recent contracts awarded to Anthropic, Google, OpenAI, and xAI by the U.S. Department of Defense[7].

These trends may undermine safety and broader governance efforts by positioning AI as crucial for economic growth and national security, thereby justifying the unsustainable industry practices (Bakhtiarifard et al., 2025; Francisco, 2023; Hendrycks et al., 2023).

**Feedback to Market Imperatives:** By failing to provide proactive governance, the AI industry is likely to follow market-driven logics (Gröger et al., 2025), as examined in the next section. This lack of oversight perpetuates environmentally destructive practices as they remain economically advantageous (Gröger et al., 2025; J. Zhao & Gómez Fariñas, 2023), intensifying competitive pressures for unsustainable growth (Hendrycks et al., 2023).

### 4.1.3. Market-Driven Imperatives

The economic incentives and competitive dynamics governing the AI industry systematically enable environmental risks. The core conflict is a misalignment between the logic of hyper-scaling, venture-backed capitalism, and the finite boundaries of ecological systems, resulting in market dynamics that reward environmental externalization and disregard for sustainable practices.

**The Growth Paradigm**: Currently, the AI sector is characterized by a growth-oriented market economy and competitive race dynamics that typically prioritize growth and rapid returns on investment (Heilinger et al., 2024; Hendrycks et al., 2023). This prevailing economic structure prioritizes rapid product deployment, creating a "growth-at-all-costs" culture that is fundamentally misaligned with ecological and planetary limits (Heilinger et al., 2024; Hendrycks et al., 2023). The AI sector's revenue is estimated to grow from approximately $200 billion

---

[5] People's Daily, April 27 2025, Accessed July 16 2025

[6] Presidential Action, January 23 2025, Accessed July 16 2025

[7] CDAO Press Release, July 14, 2025, Accessed July 16 2025

in 2023 to $1.8 trillion by 2030 (Schütze, 2024a), creating economic competition that incentivizes a business model focusing on short-term financial gains over sustainable development (Hendrycks et al., 2023).

As a consequence of this paradigm, efficiency gains of or from AI are more likely to be reinvested to expand production, develop larger models, and capture new markets rather than leading to a net reduction in resource use (Luccioni, Strubell, et al., 2025). This is intensified by AI models themselves, which tend to optimize variables that create business value (John et al., 2022), thereby embedding growth imperatives directly into algorithmic decision-making. The imperative for near-term returns steers AI development towards applications that offer high monetization potential (Read, 2025). This means that AI applications prioritizing sustainability must yield short-term financial gains to compete with less sustainable solutions offering high short-term yields (Luccioni, Strubell, et al., 2025), thereby creating a temporal misalignment, as sustainable development typically requires longer-term investment and slower development cycles.

***Competitive Dynamics and Environmental Externalization:*** The economic pressures for growth manifest in specific corporate behaviors that actively create and perpetuate systemic environmental risk (Dauvergne, 2022; Ojong, 2025). Externalizing these costs is a competitive advantage that allows companies to lower their operational expenses and enhance profitability (Dauvergne, 2022). This competition creates pressure and incentivizes other market actors to adopt similar practices (Ojong, 2025). Competition over scarce resources that power modern AI development creates additional pressure. As these resources become scarcer and more expensive, companies face increasing pressure to shift extraction and processing to regions with weaker environmental protection (Dauvergne, 2022). The importance of the "first-mover advantage" further drives rapid development, encouraging deployment of AI systems before their environmental risks are fully understood (Hendrycks et al., 2023). Companies that delay deployment to conduct environmental harm assessments may risk losing their market position to competitors, creating a "race to the bottom" that undermines safety and long-term governance (Hendrycks et al., 2023). These dynamics are also characterized in our case studies. One expert described the dominant return on investment considerations driving AI adoption in the recycling industry: "Some [customers] are purely business-based. They compare our solution to the pre-existing human solution, labor cost, and so on. And then if that payback time for the investment is less than two years or something, they move forward." These considerations indicate how, in the recycling industry, AI is not being adopted to improve environmental considerations but rather to make operations more cost-effective.

***Feedback to Opacity:*** Market pressures directly incentivize technological opacity (Brevini, 2020). Companies have strong economic reasons to resist transparency about their environmental impacts, as such information could trigger regulatory responses, consumer backlash, or competitive disadvantages (Dauvergne, 2021; Luccioni, Gamazaychikov, et al., 2025). This economic logic reinforces the technological opacity examined in the final section on structural conditions.

#### 4.1.4. Technical Opacity

The inherent technical characteristics of many AI systems, in particular their complexity, inscrutability, and unpredictability, create structural conditions that enable and conceal systemic environmental risks. This opacity operates not merely as a technical limitation but as a functional barrier to environmental governance and accountability by hindering independent assessment and auditing, and attributing responsibility.

***Inherent and Strategic Opacity:*** AI's opacity has two dimensions. The first is inherent opacity: many modern ML models, and as a consequence, the majority of modern AI systems, are known to be hard to predict or understand due to their complexity and probabilistic nature. Therefore, they are often treated as black box systems where the relationship between their input and output is unknown or difficult to explain (Z. Chen et al., 2023). The opacity of AI systems is a core inhibitor of environmental oversight and auditing, fundamentally challenging effective environmental practices (Robbins & van Wynsberghe, 2022; J. Zhao & Gómez Fariñas, 2023). The second dimension is strategic opacity, where corporations intentionally withhold information about model architecture, training data, and environmental impacts, citing intellectual property. An expert from our case study on AI in oil and gas said, "It's relatively difficult, I think, to understand what is actually being applied in reality. […] you really only have access to promotional videos and the occasional interview-based report. But you can't really dig deeper because it involves a lot of proprietary information," describing deliberate barriers to transparency. This strategic opacity increases the difficulty of determining which factors influence the decision-making process. Crucially, this entails that the opacity allows for the propagation and embedding of explicit or implicit biases, such as rapid growth or narrow efficiency gains, that might have environmental risks associated with them (Gröger et al., 2025).

***Unpredictability and Risk Assessment Failure:*** Another consequence of the probabilistic and complex design of many modern AI systems is their unpredictability. Foreseeing potential harms arising from interactions with complex systems, such as the environment, is difficult (da Costa et al., 2024; Wisakanto et al., 2025). These characteristics challenge traditional approaches to environmental risk assessment that rely on cause-and-effect relationships (Wisakanto et al., 2025). When AI systems interact with complex environmental systems, the potential for unintended consequences increases significantly (Galaz et al., 2021; Kaack et al., 2022), but current methods cannot reliably predict these interactions (Wisakanto et al., 2025).

The uncertainty of AI development itself compounds this. With rapid progress in the field of AI, systems are regularly updated and new systems are deployed, making proactively identifying and governing harms more challenging (Hendrycks et al., 2023; Ojanen, 2025). The lack of explainability also fosters emergent behaviors (Wisakanto et al., 2025). Due to the "black-box" paradigm, these emergent behaviors, particularly in complex multi-agent systems, become more challenging to identify and contend with (Wisakanto et al., 2025).

***Feedback to Power Concentration***: Opacity reinforces power concentration by making AI systems more dependent on their original developers for interpretation and modification (Brevini, 2020; Domínguez Hernández et al., 2024). This technical dependency strengthens the positions of dominant actors, making it more difficult to challenge environmentally harmful practices and completing our exemplary cycle of reinforcing drivers.

### 4.1.5. The Reinforcing Cycle

The conditions introduced in this chapter create a self-perpetuating system where each strengthens the others. Power concentration shapes governance to favor market-driven interests and resist environmental regulation. Governance gaps create permissive conditions for market-driven environmental externalization. Market pressures incentivize opacity to avoid accountability. Opacity enables power concentration by making systems dependent on their creators, while shielding harmful practices from scrutiny. The feedback mechanisms outlined in this section are by no means complete, and many more paths of self-perpetuation can be drawn and elaborated upon.

However, understanding this reinforcing cycle indicates why isolated interventions, such as voluntary corporate commitments or technical efficiency improvements, fail to address systemic environmental risks.

**4.2. RISK AMPLIFICATION MECHANISMS**

In this section, we consider the mechanisms by which systemic environmental risks of AI propagate, embed themselves, and are amplified. We posit that the structural conditions activate and enable these specific mechanisms through which AI systems amplify environmental risk. These mechanisms represent pathways by which structural conditions are translated into observable environmental consequences.

4.2.1. Rebound Effects

One of the most empirically supported mechanisms is the rebound effect, also known as Jevons' Paradox (Cellard et al., 2025; Z. Chen et al., 2023; Dauvergne, 2022; Ertel & Bonenberger, 2025; Galaz et al., 2021; Gröger et al., 2025; John et al., 2022; Kaack et al., 2022; Luccioni, Strubell, et al., 2025; Nordgren, 2022; Pachot & Patissier, 2023; Q. Wang et al., 2024; Wright et al., 2025). It was first described by the economist William Jevons (1865), who observed that the development of more efficient steam engines led to an increase in demand for coal. Thus, rebound effects describe instances where efficiency gains lead to a rise in demand and consumption, thereby exacerbating the environmental impact. Rebound effects are significantly influenced by the market-driven imperatives governing the AI industry (Luccioni, Strubell, et al., 2025). Efficiency gains achieved, be it in AI or through AI, are systematically reinvested to expand markets, increase production, and stimulate new demand. This economic logic drives aggregate resource consumption, leading to net negative impacts and rebound effects, which are systemic environmental risks of AI.

***Direct Rebound Effects*** occur when increased efficiency of an AI system leads to more frequent or extensive use of that system (Kaack et al., 2022; Luccioni, Strubell, et al., 2025). For example, when optimized logistics reduce fuel usage per delivery and shorten delivery times, this can encourage more frequent online orders, thereby increasing overall mileage and emissions within the logistics system (Luccioni, Strubell, et al., 2025). One expert for agricultural systems warns, "So I can see a world in which the Jevons' Paradox is supercharged by AI, specifically for agriculture. We already see that with access to high-quality seeds or fertilizer. Governance will be necessary not only in the development and deployment of models and data but also in the application of AI across the board. Otherwise, there are some pretty substantial downside risks of overproduction in an unsustainable way."

In AI research itself, smaller models and therefore more efficient training might also induce increased experimentation and longer training runs, ultimately expanding the resource demand (Wright et al., 2025). A clear industrial example is documented by Meta, which found that despite significant efficiency gains, its "overall electricity demand for AI" continued to increase because efficiency improvements stimulated the development of "additional novel AI use cases" (Wu et al., 2022).

***Indirect Rebound Effects*** occur where AI-driven efficiency gains free up resources, such as time or capital, that are consequently spent on increased resource consumption elsewhere (Ligozat et al., 2022; Luccioni, Strubell, et al., 2025). Typical examples are that money saved by a more fuel-efficient AI-optimized vehicle or mode of transport, such as commonly AI-optimized ride-sharing apps like Uber, might be spent on air travel or other consumer products (Luccioni, Strubell, et al., 2025). A similar indirect rebound, known as the "spare time rebound effect", occurs through time savings. A robot performing household chores may save time that is then spent on other activities like consumption or travel, a phenomenon known as the spare-time rebound effect (Ertel & Bonenberger, 2025).

***Economy-wide Rebound Effects*** describe systemic changes where AI-driven innovations transform production and consumption across entire economies, leading to a net increase in resource demand (Luccioni, Strubell, et al., 2025; Nordgren, 2022). As the effects of AI go beyond the AI sector itself (Luccioni, Strubell, et al., 2025), it has the potential to increase Gross Domestic Product (GDP), which in turn can lead to increased resource consumption, such as electricity, regardless of AI's efficiency gains (Read, 2025). These economy-wide rebound effects may cause cross-sectoral ripple effects and macroeconomic ripple, rebound, and shifting effects far removed from the original intended point of intervention of the efficiency improvement (Garske et al., 2021; Luccioni, Strubell, et al., 2025).

Despite AI-driven micro-level efficiency improvements, rebound effects entail an offset by increased consumption or production, which can lead to a net negative environmental outcome and potentially displace less environmentally harmful solutions (Ertel & Bonenberger, 2025; Luccioni, Strubell, et al., 2025; Wright et al., 2025; Wu et al., 2022), potentially only delaying the crossing of planetary boundaries rather than mitigating it (Heilinger et al., 2024). Rebound effects, thus, accelerate unsustainable systems by making harmful practices cheaper and more productive (Luccioni, Strubell, et al., 2025). By enabling more affordable goods and services, these dynamics directly contribute to the Reinforcement of Unsustainable Lifestyles & Consumption that we will discuss later. Critically, corporate sustainability narratives often overlook rebound effects, instead highlighting marginal efficiency gains while the net negative impact of increased aggregate consumption remains obscured (Dauvergne, 2022). This "greenwashing" presents a partial truth that distracts from the larger problem (Ojong, 2025), reinforcing a techno-solutionist belief that will be discussed later in this chapter. Given these complex interactions, addressing rebound effects requires interventions targeted at the underlying incentive structure through regulations that are designed to translate efficiency gains into net reductions in resource use (Ertel & Bonenberger, 2025; Luccioni, Strubell, et al., 2025).

4.2.2. Path Dependencies

Path dependencies describe how past decisions can disproportionately influence future choices, e.g., by increasing the cost of one future choice over another. Lock-in represents a stricter form of path dependency, where a past choice effectively renders alternative future choices inaccessible (Ojanen, 2025). This phenomenon is particularly relevant to AI systems as they become more deeply entrenched in our infrastructure, risking harmful trajectories, path dependencies, and lock-in (Ojanen, 2025; Robbins & van Wynsberghe, 2022).

**Mechanisms of Entrenchment:** AI depends on vast physical infrastructure and complex supply chains (Domínguez Hernández et al., 2024; Falk et al., 2024; Robbins & van Wynsberghe, 2022; Valdivia, 2024). The development and establishment of these long-lived and expensive assets create significant inertia. Changing the infrastructure and supply chains, thus, becomes challenging and costly once they are established (Robbins & van Wynsberghe, 2022), constituting a form of infrastructure or material lock-in. Another form of lock-in observable in the literature is the cognitive and institutional lock-in. As a technological paradigm, such as the transformer or convolutional architecture trained on large amounts of data and specialized hardware, becomes prevalent, this trajectory influences industry standards, educational programs, and research priorities (Bender et al., 2021; Brevini, 2021; Kaack et al., 2022). This introduces a bias in both practitioners and policymakers towards the dominant high-compute paradigm, potentially disadvantaging less environmentally harmful practices (Bender et al., 2021; Heilinger et al., 2024). One expert describes these mechanisms, drawing a historical parallel on how technology can lead to entrenchment: "the main concern [is] that digital technologies or AI allow companies to remain competitive for longer. And we saw with fracking how technological development led to huge production volumes or huge reservoir enlargements, simply because you now have access to completely different resources than before." Furthermore, the current narrow regulatory focus on direct environmental impacts of AI constitutes a form of "governance lock-in". By underrepresenting the broader systemic and supply-chain risks, this approach fails to create incentives for fundamentally different, more sustainable systems, thereby reinforcing the dominance of the existing model (Cellard et al., 2025; Heilinger et al., 2024; Luccioni, Strubell, et al., 2025).

This foreclosure of future choices leads to several systemic consequences. For one, it leads to the optimization and prolonging of unsustainable systems, since AI's application typically focuses on the incremental optimization of existing, often unsustainable, industries (Dauvergne, 2022; John et al., 2022; Ojong, 2025), also strengthening the "carbon lock-in" (Robbins & van Wynsberghe, 2022). Additionally, it interplays in several ways with other components in this chapter. As discussed, it is enabled by the concentration of power and opacity, allowing centralized, proprietary, and opaque models. These path dependencies also lead to an erosion of socio-ecological resilience by imposing these systems, which constitutes a form of digital solutionism that we will cover later in this section.

#### 4.2.3. Infrastructure Integration

Integrating AI into infrastructure poses the risk of creating fragile systems in which localized disruptions can trigger unpredictable, cascading failures across previously disconnected domains. Integration is accelerated by market-driven imperatives to pursue AI's widespread adoption, compounded by a homogenization driven by power concentration. While infrastructure integration is closely related to the path dependencies, it represents a distinct mechanism. Path dependencies describe how past decisions constrain future choices through cognitive, institutional, and material lock-in. Infrastructure integration, by contrast, describes the physical and functional coupling of AI systems into critical infrastructure, creating new risk vectors through interconnection and interdependence. Path dependencies make sustainable alternatives harder to choose, while infrastructure integration makes localized failures more likely to cascade across previously disconnected domains.

***Scope of Integration:*** The integration of AI into critical infrastructure is evident across sectors essential for social and ecological stability (Galaz et al., 2021). AI is being deployed in food systems, forestry, marine resource management, and water distribution management (Adnan et al., 2024; Galaz et al., 2021), as well as infrastructure for healthcare, transport, security, and energy, including energy grids (Adnan et al., 2024; Dhiman et al., 2024; Galaz et al., 2021), and managing urban systems such as traffic flow and waste disposal (Adnan et al., 2024; Galaz et al., 2021), while "Industry 4.0" sees AI being deeply embedded in supply chains and manufacturing (Dauvergne, 2022). This increasing integration of AI into critical infrastructure creates new social-ecological-technical systems, where algorithms, human practices, and ecological processes become inextricably linked (Galaz et al., 2021; Rakova & Dobbe, 2023), creating and giving rise to new vulnerabilities and environmental risks. The infrastructure integration is illustrated by one case study expert describing the integration into the oil and gas infrastructure: "AI is applied across the whole oil-and-gas value chain, exploration, marketing, transport, refineries, and leak detection". They further observed a new physical coupling emerging from this integration: "With the data center boom, I've already seen that the first data centers want to get a direct gas connection. […] Right, that they then also become major offtakers of fossil fuels. And that there might be a nice entanglement there. So almost like a cycle, a little bit."

***Vulnerability Mechanisms:*** The tight coupling of AI with essential infrastructure can increase the overall system's fragility. A localized disruption, such as a technical failure, a cyberattack, or an algorithm failing to adapt to out-of-distribution data, can propagate rapidly, causing cascading failures (Galaz et al., 2021; Gröger et al., 2025; Tzachor et al., 2022). The concentration of models and platforms on a few actors leads to a homogenization of technical infrastructure, making entire and unconnected sectors vulnerable to correlated failures and reducing overall resilience (Robbins & van Wynsberghe, 2022). The complexity also entails that systems are vulnerable to complex network interactions that can give rise to unintended consequences (Hendrycks et al., 2023). If multiple AI systems interact with different optimizations, the emergent behaviors are challenging to predict, and through their connection to critical infrastructure, could have severe consequences (Wisakanto et al., 2025).

### 4.2.4. Trade-offs & Competing Objectives

Sustainability considerations are typically multidimensional, incorporating environmental, social, and economic aspects that often conflict with each other (Heilinger et al., 2024). However, the complexity arising from their interconnectedness frequently leads to simplified assessments (Heilinger et al., 2024). This narrowed sustainability view, for example, focusing on reducing carbon emissions, might overlook other environmental or social risks (Falk et al., 2024), leading to problems being shifted rather than resolved (Heilinger et al., 2024). The systemic risks, thus, do not arise from the existence of trade-offs themselves, but from the systematic bias in their resolution. This bias is a direct consequence of the Market-Driven Imperatives, which structurally prioritize objectives such as computational performance, speed, and profit over sustainability. Hence, environmental degradation becomes an accepted trade-off to achieve economic gains (Gröger et al., 2025; Ojong, 2025).

**Performance vs Environmental Cost:** The AI industry is dominated by the "bigger is better" or "Red AI" paradigm of development, prioritizing task performance over resource consumption (Z. Chen et al., 2023; Schwartz et al., 2020). This drive for state-of-the-art results comes at the cost of exponential scaling of compute resources (Bakhtiarifard et al., 2025). It has become the dominant path for innovation, which directly contrasts with environmental sustainability. However, evidence increasingly suggests this trade-off is constructed by industry priorities rather than technical necessity. DeepSeek, for instance, achieved performance competitive with models in its class while using a fraction of the computational resources (Bakhtiarifard et al., 2025), demonstrating that the "bigger is better" assumption underlying Red AI reflects strategic choices about what to optimize rather than fundamental technical constraints. The opacity aids this prioritization of the AI development process, concealing the actual environmental cost of the often marginal performance gains (Chakraborty, 2024; Falk et al., 2024; Schwartz et al., 2020).

**Single-Objective Optimization:** AI systems are often designed to optimize for a single objective. This creates a form of competing objectives, where economically driven goals conflict with environmental considerations. A clear example of these competing objectives is the use of AI in the oil industry, where AI systems are deployed for economic optimization, accelerating resource discovery and extraction (Kelly, 2022). The systemic environmental risk arises from omitting environmental concerns, such as deforestation, from the optimization function. Additionally, an AI system designed to optimize a supply chain, minimizing financial cost, might select more carbon-intensive shipping routes over rail freight. One expert on waste-robotics made the engineered trade-offs explicit: "And this is something that we optimize depending on the customer needs, case by case. For some customers, they want to maximize the throughput of their plant with some constraints on what the recovery of the recyclables needs to be and what the purity needs to be. For some other customers, they want to maximize how much recyclables they recover or minimize the amount of remaining contaminants." The material stakes of these choices are real, as another expert explained: "if you are prioritizing throughput and don't have the required accuracy, then the output of that sorted recycling pile will either lead to a less valuable recycled item or less quality recycled item. […] The product quality decreases, so you are actually getting less environmental benefit from this recycling process. You might not be able to recover certain items at all." This case illustrates that depending on performance choices, such as prioritizing sorting speed over accuracy, can lead to negative environmental outcomes.

**The Pharmakon Framing:** These mechanisms are rationalized by considering AI as a pharmakon, both a poison and a remedy to the environment. This "dual role" dilemma of AI creates a moral hazard, where the promise of future AI-based solutions to climate change is used to justify current resource-intensive practices, serving to delay and distract from more immediate, non-AI-based solutions (Schütze, 2024b). The systemic bias in resolving these trade-offs is not only a strategic priority, but it also becomes a form of algorithmic bias through competing objectives, which we will explore in the following section.

4.2.5. Algorithmic Bias

Biases in AI systems, often originating from biased data, embedded values, and unrepresentative development teams, function as a mechanism for propagating systemic environmental risk. These biases originate from several sources, often leading to the unjust distribution of environmental burdens (Zhuk, 2023).

**Data-Related Biases:** It is well-documented that poor, limited, or biased training data can lead to algorithmic bias, resulting in environmental harm (Galaz et al., 2021; Zhuk, 2023). Galaz et al. (2021) hypothesize that this might affect precision agriculture, where systems trained on insufficient data might fail to provide adequate recommendations, and thus pose a risk to ecosystems and yields. They further introduce the transfer context bias, where AI systems designed for one context might fail to generalize to others. They argue that models trained on data from large, data-rich industrial farms are likely to fail in the context of small farms, potentially in different climates, resulting in potentially damaging outcomes (Galaz et al., 2021). As the access to the development of AI systems is, as discussed earlier, concentrated mainly in HICs, a lack of geo-diversity is also highlighted as a crucial source of bias. This implies that datasets may not fully represent certain regions, and therefore, AI systems may underrepresent specific regional contexts (McGovern et al., 2022). Datasets collected with specific parameters may inadvertently overrepresent certain environmental conditions, leading to skewed results and unfair outcomes, which often affect marginalized groups (Zhuk, 2023). These biases are also illustrated in our case studies. One waste-robotics expert reported direct non-transferability, discussing how their team trained models on data from different waste-management facilities. "And the model that is trained on one facility was not successful in the other facility, even though the geographical distance is not that much." The other expert from the case study concurs that their company changed their strategy due to models not generalizing across facilities, stating: "When I joined the company, we were still trying to build one universal model that understands everything about all kinds of waste and that can be applied everywhere. And then it turned out that developing such a model was very costly in terms of effort. And also, it started to look like it might be that already on the concept level, it's not a well-defined thing."

**Value-Embedded Biases:** As Kate Crawford (2016) asserts, "Like all technologies before it, artificial intelligence will reflect the values of its creators.", highlighting the extent by which AI systems are shaped by the sociocultural biases of their developers as well as the economic system in which it is embedded (Luccioni, Strubell, et al., 2025). This is driven by a lack of diversity in AI research and governance (Debnath et al., 2023; Domínguez Hernández et al., 2024). AI systems designed with limited diversity can create harms that only become apparent when real-world impacts affect communities not considered in the training data or system design (Kelly, 2022). Training data, especially pervasive, uncurated datasets from the internet, can encode hegemonic worldviews that are detrimental to marginalized populations (Bender et al., 2021). These embedded biases and injustices can lead to the development of algorithms that perpetuate existing societal inequalities and may "hallucinate" during decision-making applications, particularly in climate-vulnerable regions with limited data justice (Debnath et al., 2023). One case study expert on food systems illustrated the environmental consequences of these biases: "Small-scale farmers, specifically in the Global South, who are more likely to engage in farming with indigenous or underutilized crops might be overlooked or the models might not apply to them quite as well." This data-related bias is a reflection of embedded values: the system is trained on data from large industrial farms because they represent the most profitable market, thus encoding an economic value that marginalizes subsistence agriculture and its associated biodiversity.

4.2.6. Information Ecosystem Effects

Systemic environmental risks arising from information ecosystem effects involve the use of AI to create, scale, and micro-target the dissemination of disinformation, which undermines the shared understanding of reality required for collective environmental action (Hendrycks et al., 2023; Uuk et al., 2024). It functions as a systemic environmental risk by potentially weakening the social and political will to develop and implement pro-environmental policies (Francisco, 2023), thereby locking in unsustainable systems and behaviors. This mechanism is driven by algorithms on social media platforms, which are designed for engagement and are influenced by economic incentives (Hendrycks et al., 2023; Uuk et al., 2024), as well as the concentration of power where a few platforms control the information sphere (Hendrycks et al., 2023; Sætra, 2021). The lack of governance measures to regulate platforms and their AI strategies further exacerbates the spread of misinformation and erosion of trust. AI has the potential to exacerbate misinformation and erode trust, including in relation to environmental issues (Domínguez Hernández et al., 2024; Hendrycks et al., 2023; Sander, 2025), posing a critical challenge to addressing climate change (Sander, 2025; van der Ven et al., 2024).

***Misinformation Generation and Amplification:*** Advancements in generative AI have played a critical role in misinformation by lowering the cost and increasing the feasible scale of producing false content (Sander, 2025; Weidinger et al., 2022), primarily by removing the need for human supervision (Hendrycks et al., 2023). The low-cost generation of misinformation could be used to fabricate scientific reports, expert testimonies against climate action, or other fraudulent content undermining environmental efforts. AI-algorithms on social media platforms, often designed to maximize engagement, can lead to the dissemination of polarizing and false content (Sætra, 2021; Sander, 2025). Sander (2025) argues that these mechanisms can be exploited to micro-target specific demographics with climate denial narratives as well as "greenwashing" advertisements. Haider et al. (2022) provide the example of a U.S. Amazon search for books on climate change, consistently providing a book "commonly described as promoting a denialist agenda" as editors pick over several months. They further note that selecting the "Kindle Unlimited eligible" criterion results in the first three books denying the existence of anthropocentric climate change. Similarly to ranking books highly that do not represent scientific consensus, AI-driven bot networks can be deployed to simulate "majority opinions", potentially making pro-environmental measures more challenging to advocate for (Weidinger et al., 2022).

***Algorithmic Amplification:*** van der Ven et al. (2024) show that AI-based chatbots can provide climate-related misinformation without explicit intent. They identify several problematic system behaviors. Chatbots often provide non-existent sources or obscure them by bundling sources, creating an illusion of authority that is further exacerbated by an authoritative tone. Additionally, the study reveals that chatbots tend to assign accountability to investors or the finance sector less frequently than to government or individual responsibility. Additionally, they tended to emphasize incremental solutions to climate change, underemphasizing systemic causes and investor accountability as causes of climate change and reinforcing the "status quo".

***Trust Erosion:*** The primary systemic environmental risks posed by this mechanism are the erosion of trust in scientific institutions, democratic processes, and established environmental knowledge (Domínguez Hernández et al., 2024; Weidinger et al., 2022). By undermining scientific consensus, misinformation can also be used to delegitimize and challenge governance measures (Francisco, 2023). Additionally, the concentration of power enables large technology firms to utilize their platforms to exert an outsized influence over the climate change discourse (Sander, 2025), potentially leading to incremental solutions that are misaligned with the urgency of the climate crisis (van der Ven et al., 2024). This erosion of public trust and misrepresentation of scientific consensus create the necessary conditions for overly simplistic narratives to establish themselves. It is within this eroded informational landscape that the mechanism of Digital Solutionism thrives, offering the promise that complex environmental problems can be solved by technology alone, a topic to which we now turn.

### 4.2.7. Digital Solutionism
Digital solutionism is the belief that complex societal and environmental problems can be resolved primarily through technological innovation, particularly AI, without fundamentally altering underlying economic structures, consumption patterns, or political power dynamics. "AI for Good" initiatives, while acknowledging the potential benefits, can be a manifestation of this ideology, often framing AI as a neutral tool for positive change (Galaz et al., 2021). We will examine how this belief, as critics argue, can obscure negative environmental consequences, misdirect resources, and hinder the necessary systemic changes for sustainability (Dauvergne, 2022; Nost & Colven, 2022). The dual framing of AI as both an environmental solution and a risk is used to resolve tension by quantifying environmental costs to ensure that positive impacts outweigh the negative (Cellard et al., 2025).

***Consequences of Solutionist Framing:*** "AI for Good" initiatives often position AI as a win-win solution, emphasizing efficiency gains and the capacity to optimize processes (Dauvergne, 2022). However, this focus on quantifiable metrics obscures the widely accepted notion that the systemic environmental risks posed by AI are more complex to quantify and can be underestimated when attention centers on easily measurable efficiency metrics (Galaz et al., 2021; Gröger et al., 2025; Luccioni, Strubell, et al., 2025). This functions, intentionally or unintentionally, as a form of greenwashing and reputation management, allowing corporations to cultivate an environmentally responsible image while their core business practices continue to cause harm (Dauvergne, 2022). One case-study expert highlights the current paradigm of digital solutionism, stating "Everything has to be AI now" and adding that "we should always be careful not to be shooting at sparrows with cannons. And just because everything has to be AI now, if the same query requires ten times as much energy as a normal statistical query or a simple correlation query, then I don't necessarily have to do it." The "digital solutionist" framing, thus, reduces complex societal and environmental problems to technical ones that can be solved using AI (Cellard et al., 2025; Schütze, 2024b), while overlooking the underlying political and economic structures (Cellard et al., 2025; Dauvergne, 2022). By privileging computational solutions, there is a risk of adopting incremental approaches rooted in past data (van der Ven et al., 2024), as well as devaluing local knowledge in sustainable resource management (Lehuedé, 2025; Nost & Colven, 2022).

***Systemic Environmental Consequences:*** First, by presenting AI as a panacea, digital solutionism can divert investment, research focus, and policy attention away from other solutions (Heilinger et al., 2024; Luccioni, Strubell, et al., 2025; Schütze, 2024b). This can hinder other climate actions, such as reducing production and consumption, by reinforcing the idea that fundamental structural changes are not necessary, as incremental technological progress is presented as the solution (Heilinger et al., 2024; Schütze, 2024a). Second, the narrative reinforces unsustainable systems. AI applications positioned as solutions often serve to optimize and thereby entrench existing unsustainable systems, such as more efficient fossil fuel extraction, rather than enabling sustainable alternatives (Dauvergne, 2022). This reinforcement of existing systems can delay a sustainable transition and can entrench AI as an infrastructure, making sustainable change more difficult (Dauvergne, 2022; Robbins & van Wynsberghe, 2022). The systemic environmental risks engendered by this entrenchment of AI in critical and complex infrastructure are what we will consider in the next section.

**4.3. MANIFESTED IMPACTS**

To conclude our framework, we examine how the structural conditions and risk amplification mechanisms culminate in the manifestation of societal and environmental impacts. Crucially, these manifested harms are not inert endpoints. They themselves, like the other aspects of our framework, can reinforce other structural conditions, amplification mechanisms, or manifested impacts, perpetuating the cycle of systemic risks.

4.3.1. Material Harms

***Environmental Externalities:*** The literature suggests that AI's environmental impact exhibits geographical asymmetries, mirroring broader patterns of environmental injustice. An example of geographic asymmetries can be seen in the physical infrastructure of AI. The requirement for REEs, water, and energy can place significant stress on communities. These practices are often localized in specific regions, often in countries with weaker environmental regulations and less political power to resist harmful activities (Domínguez Hernández et al., 2024; Falk et al., 2024). Falk et al. (2024) highlight how communities in mining regions face increased exposure to toxic substances, water contamination, and ecosystem disruption. Valdivia (2024) discusses this with a fieldwork-based study observing that AI infrastructure, such as data centers, is celebrated in rural areas while placing ecological pressure on local communities. They highlight how local communities are faced with water shortages due to the water requirements for cooling the data centers. Quantifying these harms beyond case studies, however, remains difficult, as attribution to AI or specific actors remains challenging (Falk et al., 2024).

The cognitive buffer separating AI development from environmental impact may weaken accountability mechanisms, allowing these practices to continue unaddressed (Brevini, 2020). Sampatsing et al. (2025) show through an interview study that during corporate decision making, "sustainability concerns played no significant role in AI adoption decisions and were largely overshadowed by other drivers", indicating the financial incentives structurally enabling these practices. This displacement is not unique to AI but may be particularly pronounced given the industry's regional concentration, rapid scaling, and reliance on globally distributed supply chains (Dauvergne, 2022; Falk et al., 2024; Murdock, 2025).

*Impact on Non-Human Life:* There is an increasing risk that AI may influence land use patterns in a way that may affect ecosystems and non-human life (Domínguez Hernández et al., 2024; Galaz et al., 2021; Zhuk, 2023). Applications, such as precision farming, while potentially having positive effects, also may incentivize land use changes, including the expansion of monoculture production at the expense of agricultural landscape and ecosystem diversity (Galaz et al., 2021). The net impact of deploying these systems on the environment has yet to be widely considered, often being poorly elaborated upon in prominent reports (Galaz et al., 2021; Luccioni, Strubell, et al., 2025). Additionally, AI is driving large infrastructure projects, such as large-scale distribution infrastructure for AI-optimized logistics or data centers that power AI itself, which pose environmental harm by fragmenting habitats and encroaching on sensitive ecosystems (Zhuk, 2023). AI's increasing use in extractive industries poses an additional risk by enabling faster exploration and making new sites financially viable (Gröger et al., 2025; Sander, 2025). However, it also drives new extraction practices, such as deep-sea mining, which might have detrimental effects on yet undisturbed ecosystems (Zhuk, 2023). A waste-robotics expert illustrates how AI can cause environmental degradation if not carefully designed, audited, and controlled: "So basically when it goes to composting, taking out plastics and metals and whatnot. The end product should be something you can place anywhere, simply spread in nature. And if there was no control, then it could be that you're putting plastics and whatnot to the sides of roads under the grass."

AI is being further integrated into ecosystem management, including fisheries and forest management, as well as wildlife monitoring (Adnan et al., 2024; Galaz et al., 2021). These systems introduce a new category of risk into complex adaptive systems, potentially leading to new and cascading failures with far-reaching environmental and social consequences (Galaz et al., 2021).

Thus, while the evidence for large-scale ecosystem disruptions remains unquantified, the observed trends and potential impacts warrant careful monitoring and again call for transparency and regulatory oversight that is currently underdeveloped.

4.3.2. Epistemic & Cultural Harms

**Marginalization of Diverse Epistemologies:** AI systems trained on large-scale, quantitative datasets may systematically undervalue environmental knowledge, such as local management practices, that are more difficult to digitize or quantify, despite their proven effectiveness over long periods (Francisco, 2023; Galaz et al., 2021). This may displace epistemologies by providing recommendations that contradict local knowledge without acknowledging the conflict, centralizing knowledge to a few dominant AI systems, shifting attention and investment from local practices to technological solutions, and increasingly positioning data-based decision-making processes over community-based knowledge (Galaz et al., 2021; Gröger et al., 2025; Kaack et al., 2022; Nost & Colven, 2022). This dynamic is illustrated in agriculture, where one expert acknowledged "you have the risk of individuals unlearning traditional techniques, not using indigenous crops, and then becoming overly reliant on a technology." They continue that "you might encounter a situation in which indigenous knowledge of wild plants is starting to be eroded because people become reliant on AI tools that disregard them", highlighting the epistemic environmental harm that AI may cause.

This marginalization becomes a systemic environmental risk when it leads to the widespread adoption of technologies that are less effective at achieving environmental goals, e.g., high-purity recycling, than the human-centric systems they replace, all while creating brittle, technology-dependent systems that lack local resilience. This process also results in capacity and dependency effects, where the displacement of local knowledge and, therefore, a reliance on algorithmic decision-making creates a technological dependency, reducing resilience if systems fail or become unavailable (Galaz et al., 2021; Tzachor et al., 2022).

**Reinforcement of Unsustainable Lifestyles:** AI is increasingly employed to modify consumption patterns. AI-powered recommendation systems, targeted advertising, and personalized content are just a few examples of the use cases where AI has become ubiquitous (Dauvergne, 2022; Gröger et al., 2025; Luccioni, Strubell, et al., 2025). Companies not relying on them face a competitive disadvantage. These systems are inherently designed to change user behavior more effectively, thereby increasing consumption and posing environmental risks (Gröger et al., 2025). Moreover, rebound effects, such as the time rebound effects, can also lead to more unsustainable lifestyle choices (Luccioni, Strubell, et al., 2025).

This is compounded by the influence AI can have on the information sphere. By influencing how individuals interact with information in general, AI also affects the environmental information through search algorithms, content recommendations, and automated content generation (Sander, 2025; van der Ven et al., 2024). This control over information can negatively affect the public's perception of environmental issues and weaken the demand for sustainable change (Sander, 2025; van der Ven et al., 2024). Additionally, by enabling the scalability of misinformation through generative AI, unsustainable lifestyles can be further entrenched, reinforcing echo chambers and perceived consensus, delegitimizing climate action, and justifying environmentally harmful behavior (Domínguez Hernández et al., 2024; Sander, 2025).

4.3.3. System-Level Consequences

The widespread adoption of AI systems can lead to system-level consequences with the potential to alter the stability of socio-ecological systems. They represent the culmination of the risks analyzed thus far and pose the risk of large-scale, unpredictable disruptions to essential systems.

**Cascading Failures:** Cascading failures are identified as a system-level consequence of the diffusion of AI into many, potentially interconnected, systems. A primary system-level consequence is the increased likelihood of cascading failures (Galaz et al., 2021; Wisakanto et al., 2025). AI systems function as hyper-connectors, creating complex systems of previously disconnected domains, increasing the complexity of critical infrastructure (Galaz et al., 2021). These novel connections create new risk vectors leading to cascading failures, where technical faults, malicious intent, or any other of the previously discussed risk amplification mechanisms can propagate rapidly through these connections, triggering chain reactions and cascading failures across previously disconnected domains (Galaz et al., 2021; Uuk et al., 2024). As AI is deployed in critical sectors, it adds new nodes and connections to an existing network, increasing its connectivity and complexity (Galaz et al., 2021). This increased connectivity, while often aimed at boosting efficiency, can create new pathways for disruptions to spread (Galaz et al., 2021). For example, the large-scale deployment of AI in agriculture, while promising gains in productivity, could simultaneously "leave growers and agrifood supply chains open to cascading accidents and cyberattacks" (Tzachor et al., 2022). An initial failure could propagate rapidly through an interconnected food system reliant on AI for irrigation, pest control, or logistics. Several distinct pathways for AI-induced cascading failures are identified. One major pathway is through bias propagation within integrated AI decision-making systems. An error or bias in a single AI model can produce "ripple effects" as its biased output is used as input for other models, leading to a cascade of errors across the entire system (Tripathi et al., 2024). A second pathway involves interactions between multiple AI systems that generate "novel failure pathways or collective behaviors" that would not arise from a single agent in isolation (Wisakanto et al., 2025).

The interconnectedness and potential novelty of these risks make assessing and attributing them more challenging, to the extent that existing risk assessment models fail to capture them (Wisakanto et al., 2025). Additionally, the technical homogenization worsens cascading failures. The growing dependence on a few dominant foundation models controlled by a handful of corporations predominantly from HICs implies that entire sectors might rely on similar or identical AI systems, creating a single point of failure and making them vulnerable to shared disruptions (Ojanen, 2025; Uuk et al., 2024).

***Systemic Erosion of Resilience:*** Next to cascading failures, AI-driven optimization has the potential to systematically erode the resilience of essential socio-ecological systems. The resilience of these systems often relies on diversity and redundancy, typically at odds with short-term efficiency optimization (Galaz et al., 2021). An expert on food and water security explained this fundamental trade-off of AI in agriculture: "If you become over-reliant on any one technique or technology, you reduce your own inherent resilience", adding, "For us, resilience is bolstered by diversification […] diversification of techniques […] diversification of crops […] ability to quickly access a diversity of tools."

The erosion of ecological resilience begins with the extractive nature of the AI industry. Through its resource demand for hardware production, data center construction and operation, as well as waste generation, the AI industry exerts pressure on ecological resilience (Adnan et al., 2024; Dacon, 2023; Falk et al., 2024). The resulting environmental degradation, such as water depletion, habitat fragmentation, or encroachment on sensitive habitats, reduces the capacity of natural systems to absorb shocks and sustain essential ecosystem services, thereby eroding their inherent resilience (Rakova & Dobbe, 2023).

The erosion of resilience extends beyond ecological resilience. Path dependencies are an example of this. The entrenchment of unsustainable practices, institutional and behavioral norms, creates path dependencies that reduce societies' resilience and adaptability by creating rigid and "locked-in" infrastructure systems (Robbins & van Wynsberghe, 2022).

Finally, the erosion of resilience extends into social and socio-ecological systems. The deployment of AI, as discussed, can lead to the loss of local knowledge, which is often essential for the resilience of communities and the ecosystems they manage (Francisco, 2023; Galaz et al., 2021). The potential of AI systems to exacerbate social inequalities can also reduce resilience in vulnerable communities by undermining their well-being and ways of life, notably when their concerns are excluded from dominant AI ethics discourses that justify localized harm for a perceived greater good (Lehuedé, 2025; McGovern et al., 2022). The result of these effects is a weakening of social and ecological adaptability required to navigate environmental challenges and climate change.

4.3.4. Summary

Throughout this chapter, we considered how the systemic environmental risks of AI emerge as a result of the current AI paradigm. They are a group of material, epistemic, and system-level harms that are the consequences of the system's underlying architecture. As many of these risks are novel or emergent, lack attribution and transparency, and current risk assessment methods are limited in their applicability to these problems, quantifying, foreseeing, or addressing these problems remains difficult. Thus, this chapter aims to characterize and exemplify these pathways, providing a structure for analyzing the systemic environmental risks of AI. As AI systems become increasingly ubiquitous in our society and industry, it is critical to consider their potential harms, emergent risks, and their impact on climate change, ensuring responsible development.

Crucially, our proposed framework is not intended as a static list of harms, but rather to characterize different harm pathways within a system of interconnected risks that exemplifies the systemic nature of these risks. To illustrate this, we consider the following illustrative feedback loops within our framework of systemic environmental risks of AI:

***The "Solutionism"-Dependency Loop:*** Starting with the ideology of "Digital Solutionism", AI systems are promoted as objective, data-based solutions for complex environmental problems. As a consequence, they are increasingly introduced into local contexts, such as waste management, agriculture, or fisheries. Consequently, context-specific knowledge, such as seasonal and regional waste specifics, soil conditions, fish stocks, or catching quotas, atrophies as more emphasis is placed on data-driven insights. This "de-skilling" creates a dependency on proprietary AI systems to manage local resources, strengthening concentration of power and validating the initial "solutionist" claim, while eroding local resilience and autonomy.

***The Consumption-Reinvestment Loop:*** As market-driven imperatives dominate the development of AI systems, their primary goal is often growth and profit. Consequently, companies typically deploy systems, such as personalized recommender systems or targeted advertising, designed to modify user behavior and increase sales. This leads to a reinforcement of unsustainable lifestyles through increased consumption, thereby accelerating resource demand and ultimately contributing to ecosystem degradation. The resulting increased profits are then reinvested in the development of more sophisticated AI models, strengthening the market imperatives and creating a self-reinforcing cycle of increased consumption and environmental impact.

***The Misinformation-Governance-Gap Loop:*** With the use of AI models to optimize engagement and generative AI to lower the cost of targeted content, actors can scale up the creation and dissemination of misinformation. This can be used to delegitimize climate science and political measures, creating the false impression of public dissent and weakening the demand for environmental regulation. As a consequence, public consensus and political initiatives, both prerequisites for strong environmental policy, can be weakened. This results in governance gaps where regulatory oversight of the technology sector and environmental policy are delayed or abandoned. The deregulation directly benefits the actors who use AI to spread misinformation, perpetuating the cycle.

These dynamics, such as the emergent nature of these risks or the reinforcing feedback loops, highlight why singular interventions often fail. To establish effective policy, these systemic risks must be considered in their entirety, and holistic risk assessment becomes a necessity.

**4.4. CASE STUDY: AGRICULTURE & BIODIVERSITY**
In this case study, we examine the systemic environmental risks of AI's deployment in agriculture and biodiversity monitoring. We interviewed a lead national biodiversity data infrastructure expert (Expert A) and a global food and water security policy expert (Expert B). Both experts outline potential systemic environmental risks associated with AI. They consider that AI's current trajectory has the potential to exacerbate global inequalities, erode local knowledge, and create new systemic vulnerabilities. We employ our framework to consider these potential systemic environmental risks arising from the use of AI in agriculture and biodiversity monitoring.

4.4.1. Structural Conditions
**Power Concentration & Asymmetries:** Both experts discuss how power is concentrated in a few private and public actors in the industry. Expert B highlights a growing global divide, noting the "big risk is that all the data and the money for training is going to already wealthy countries, relatively wealthy farmers," which will cause the "AI gap, will grow even wider between the Global North and Global South". They also notice that "you can see that widening even within a country and then a conglomeration within countries that mirrors the larger global effect." This dynamic is mirrored by private sector dominance in developing these technologies. A lot of innovation in computer vision, drone technology, and digital twins is "being done in the private sector," leading Expert B to "see a world in which it is not a global public good".

Expert A describes how crucial environmental data is held in inaccessible private silos. In Germany, biodiversity data from environmental impact assessments is collected by "private firms" and "is usually not made public. It only goes to the client". This privatization of data is a significant barrier to public and scientific oversight. Expert A further notes that "Authorities have a lot of data, but they don't make some of it available electronically."

**Governance:** The governance of AI in this sector is characterized by data fragmentation and skepticism about the efficacy of evidence-based decision-making. Expert B points out that "there isn't like a hugely centralized pool of data that can be used to train models in an equitable way," which is a prerequisite for fair global deployment.

Both experts express concern over regulatory will and consistency. Expert A is critical of "Evidence-based decisions in politics," stating, "I don't know if I still believe in that". They cite the premature shutdown of a promising automated biodiversity monitoring project by Germany's BMBF as an example of inconsistent government support. Both experts also warn of ideologically driven moves to stop collecting crucial environmental data. Expert B notes that certain data is no longer collected at the USDA, which is "not great".

**Market-Driven Imperatives:** Both experts express that economic incentives primarily drive AI adoption in agriculture. Expert B notes that while being "ecologically minded" can be an incentive when it aligns with profit, "The ultimate goal, I think, mostly would be to be profitable […] I wouldn't say that most people have sustainability over profit currently".

The market imperatives are also a challenge to data sharing, which would be crucial in biodiversity monitoring, for instance. Expert A gives the example of fisheries data, where a "high commercial value" and concerns over EU fishing quotas lead countries and fishermen to not "make everything so public". Expert B further elaborates that the competitive dynamic further accelerates inequalities, as the gap between technology adopters and non-adopters can grow "so quickly as to be sort of insurmountable".

**Technical Opacity:** Both inherent and strategic opacity obscure a full understanding of AI's impact. Expert A emphasizes that crucial background information on biodiversity "exists in many places, but it is often in silos, in individual databases, and thus not accessible". They also point to a politically motivated opacity, where there is a "systematic attempt not to collect certain information anymore so that politics can later say, 'Well, I didn't know' or 'There was no data for it.'". This is compounded by the inherent opacity of the models themselves, which Expert B notes are often so "incredibly complex. Often those that make them don't understand what's actually happening".

### 4.4.2. Risk Amplification Mechanisms

**Rebound Effects**: Expert B explicitly identifies the risk of a rebound effect in agriculture. They state, "I can see a world in which the Jevons' Paradox is sort of supercharged by AI, specifically for agriculture". In this scenario, efficiency gains do not lead to a net environmental benefit but instead encourage "overproduction in an unsustainable way"

**Path Dependencies:** Expert B warns that AI could entrench existing unsustainable agricultural models and create new forms of technological lock-in. They fear that if not carefully managed, AI will "reinforce sort of the ecological degradation, the agricultural biodiversity erosion that we've seen for the last, I don't know, two centuries". Simultaneously, it "also potentially locks farmers into being reliant on this tool, which would be AI," creating a vulnerability where an over-reliance can have "detrimental effects once that's taken away".

**Infrastructure Integration:** The primary channel for AI integration in Global North agriculture is through existing mechanized infrastructure. Expert B explains that "mechanization is the biggest thing. I think John Deere covers the earth like every three days with their machines". Applying AI tools like computer vision to this existing infrastructure provides a "huge, huge sort of lever for reducing labor, potentially reducing costs, and certainly increasing the efficiency of agriculture".

**Trade-offs & Competing Objectives:** The deployment of AI involves complex trade-offs between performance, resource use, and social equity. Expert A warns against a mindset where "everything has to be AI now," cautioning against "shooting at sparrows with cannons". They argue that if an AI query requires "ten times as much energy as a normal statistical query" for little added value, "then I don't necessarily have to do it". Expert B frames the issue in terms of social trade-offs, such as "personal information, privacy, versus democratization, access to information and data" and notes that centralization, while useful, carries the "risk of overlooking marginalized communities".

**Algorithmic Bias:** Both experts highlight potential biases in their fields. Algorithmic bias from unrepresentative datasets poses a risk to both equity and environmental outcomes. Expert B identifies a "huge risk" that models are being trained with data that "aren't applicable globally, and specifically locally around the world". This particularly disadvantages "Small-scale farmers, specifically in the Global South, who are more likely to engage in farming with indigenous or underutilized crops," as those crops may be "overlooked" by the models.

Expert A describes how data collection methods introduce bias. Citizen scientists mapping biodiversity tend to do so along accessible roads and paths, leading to a bias where bird sightings correlate with roads not because birds fly there, but "rather where the people were who were watching and counting them". This results in "many white spots on the map in Germany regarding biodiversity". They also state that there are temporal biases, noting that "research projects are always very time-limited. That is, they begin and end. During that time, very good data is certainly collected, but always only for very small time spans." These data biases pose a threat by potentially introducing algorithmic biases if not carefully accounted for.

**Digital Solutionism & Framing:** Both experts are critical of the techno-solutionist framing surrounding AI. Expert B notes the common but false dichotomy that AI will "either save us all or it's going to kill us all," concluding, "It's going to be in between. I don't think it's going to save us all". Expert A criticizes the assumption "that the data is just somehow there, and now we're all doing AI". They state "Everything has to be AI now." Expert A uses the metaphor of an "inverted pyramid […] standing on its tip, with a small data foundation at the bottom and a huge AI head built on top", to illustrate the danger of building complex AI systems on an inadequate data infrastructure.

### 4.4.3. Manifested Harms and Impacts

**Material Harms:** The most immediate material harms stem from faulty AI models or the malicious use of accurate data. Expert B warns of the danger of AI systems "producing bad advice through AI modeling," which could be "detrimental" to crop yields and farmer livelihoods if the model's responses "are false or have hallucinations or aren't applicable locally".

A different material harm arises from making sensitive biodiversity data public. Expert A describes the conflict between open science and protecting vulnerable species, warning, "I don't want to have the last unicorn that was sighted with its GPS position on the map, because then someone will surely be found who wants to shoot this unicorn". They conclude that to prevent this, data on endangered species must be obscured.

**Epistemic & Cultural Harms:** A risk identified by the experts is the erosion of local and traditional knowledge systems. Expert B warns of the "risk of individuals unlearning traditional techniques" and of "indigenous knowledge of wild plants […] starting to be eroded because people become reliant on AI tools that disregard them". This is mirrored by a loss of formal scientific expertise. Expert A notes that deep taxonomic knowledge, such as that of a "spider researcher," is "in the truest sense of the word, dying out," and the goal is now to "transfer the knowledge that is now held by experts into the AI".

System-Level Consequences: At a system level, the experts identify several consequences that could destabilize food systems and undermine environmental governance. Expert A points to a breakdown in the evidence-based policy process, where there is a "systematic attempt not to collect certain information anymore so that politics can later say, 'Well, I didn't know'".

The over-reliance on homogenized, AI-driven agricultural practices threatens to "reinforce […] the ecological degradation, the agricultural biodiversity erosion". Expert B further argues that becoming "over-reliant on any one technique or technology" reduces a system's "own inherent resilience," as resilience is "bolstered by diversification". This creates a new vector for malicious attacks. Expert B warns that a "bad actor" could subtly "tweak" AI models to produce bad advice, creating a "vector through which people that will gain something by agriculture collapsing in one place to engage in that type of behavior".

**4.5. CASE STUDY: OIL & GAS**
In this case study, we examine the systemic environmental risks arising from AI's deployment in the oil and gas industry. The case study is based on two expert interviews. We interviewed an academic petroleum engineer (Expert A) and an academic expert on climate technology policy (Expert B). Both experts outline that AI is being used throughout the value chain of the industry. Both acknowledge that AI has environmental benefits, such as methane leak detection, but conclude that the dominant pattern is production optimization and extraction acceleration, which both function to strengthen the industry's competitiveness. Partnerships between major technology companies and fossil firms shape the channel through which AI enters the energy system. Expert B points to Microsoft as "quite well-known for having a relatively large number of contracts" in oil and gas and links this to potential alignment of lobbying activities with continued extraction. Throughout this case, the experts indicate how AI strengthens extraction through structural concentration, amplification mechanisms, and manifested harms. We apply our framework to the case study to illustrate the systemic environmental risks of AI in the oil and gas industry.

### 4.5.1. Structural Conditions

**Power Concentration & Asymmetries:** Both experts highlight how data centralization and Big Tech-fossil alliances concentrate decision rights and shape AI toward extraction goals rather than environmental protection. Expert A describes how, in India, geological and production data are centralized under the Directorate General of Hydrocarbons. While companies receive data access for licensed exploration blocks, "a central agency is holding all the data," making it "very difficult for academics and for scientists to work because they don't have data." This concentration limits independent research, public oversight, and the development of alternative applications that might prioritize environmental protection over extraction efficiency.

At the global scale, a new power concentration emerges from partnerships between major technology corporations and oil and gas companies. As Expert B notes, "Microsoft is quite well-known for having a relatively large number of contracts with oil and gas […]." The concentration has the potential to create political leverage: "If Big Tech also sees oil and gas as partners and has more political weight, one could imagine that they might then align their lobbying activities accordingly." The alliance between two of the world's most influential industrial sectors potentially strengthens resistance to climate regulation.

Expert A finally highlights how optimization-driven extraction affects local populations: "If an oil and gas well has been found […] The well will be drilled and everything will be done, irrespective of environmental damage, irrespective of the people. Everybody will be relocated or whatever. This is how progress is started." AI's role in the oil and gas industry, thus, also has the potential to benefit large corporations, while local communities increasingly bear the burdens.

**Governance:** In addition to the increased potential for lobbying highlighted by Expert B, both experts discuss regulatory gaps and misaligned accounting that allow AI to raise output while sidestepping climate alignment. Expert B emphasizes that robust climate policy should apply to AI, yet argues that AI-specific laws are needed to exert stricter control over direct emissions. They further contend that general climate policy alone may be insufficient to align the sector with climate goals. Expert A adds that day-to-day industry practice still gives low priority to environmental safeguards, and that AI can be used to locate and exploit regulatory loopholes unless governance keeps pace.

Both experts characterize the regulatory landscape as reactive. Expert B advocates for strict general climate policy to regulate systemic environmental risks of AI, while acknowledging "[…] normal climate policy is not sufficient." Expert A further underscores the need for regulation, stating that, although environmental concerns are improving, they are not sufficiently considered by the oil and gas industry.

Expert B also argues that relying on voluntary corporate commitments is unlikely to address systemic environmental risks. They note that "[…] the tech sector claims to want to become climate neutral, but on the other hand, services for oil and gas are not included in their scope." This omission creates an accounting gap in which "Scope 3 emissions are completely ignored, […]" despite worker campaigns within the technology industry to include them.

**Market-Driven Imperatives:** Economic incentives define the scope and performance metrics of AI in oil and gas. Across interviews, AI is characterized as a tool that serves production and cost objectives rather than decarbonization goals. The experts explain how profitability and competition define AI objectives, which channels model design toward production and cost metrics. When asked about strategic objectives for AI investment, Expert A identifies "production improvement, production optimization, reservoir management" as central goals, with health and safety as aspirational rather than primary concerns. Expert B reinforces this assessment, noting that oil and gas represent "simply a lucrative business" where technology partnerships are driven by "potential revenue" rather than environmental criteria, explicitly designed to "increase production volumes."

Both experts comment that competition drives AI adoption in the oil and gas. Expert A notes that companies try to infer rival bids, production profiles, and infrastructure characteristics, then adjust strategies accordingly. They elaborate that such dynamics give rise to aggressive data acquisition and increase exposure to integrity and confidentiality risks. Expert B underscores the market outcome of these dynamics. They express "[...] that these digital technologies or AI allow these companies to remain competitive for longer [...]". The combined effect is intensified competition on efficiency rather than on absolute emissions, which delays exit from carbon-intensive production and deepens lock-in.

**Opacity:** Across interviews, opacity is considered an enabler of systemic environmental risk. The experts detail strategic and inherent opacity that limits verification of environmental claims and weakens public oversight. Expert B explains that "[...] it's relatively difficult, I think, to understand what is actually being applied in reality, what the companies are using." They add that opacity is partially due to proprietary information and that investigations "[...] only have access to promotional videos and the occasional interview-based report [...]." Expert A further elaborates how, in India, power concentrations affect opacity, explaining that it is "very difficult for academics and for scientists to work because they don't have data."

Both experts highlight limits to prospective assessment. As Expert A puts it, "we are forecasting the 2030 data, so we do not know whether it is correct or not until we reach 2030." This uncertainty interacts with opacity, which delays verification and weakens accountability for environmental claims.

Together, these mechanisms help obscure the effects AI has on the oil and gas industry and limit auditing and regulation.

### 4.5.2. Risk Amplification Mechanisms

**Rebound Effects:** Both experts identify AI-driven efficiency, potentially leading to higher aggregate extraction, converting micro-efficiencies into macro-level emissions growth. When asked whether AI is more likely to advance the oil and gas industry's alignment with climate goals or accelerated extraction, Expert A concludes that "I think the second one is more correct than the first one [...]."

Expert B echoes this sentiment, noting that AI is used to "[...] to increase production volumes [...]". They also caution against efficiency gains in the sector, stating that "[...] any kind of efficiency increase through AI in the entire value chain definitely has rebound effects [...]". Together, the interviews suggest that AI-driven efficiency gains are more likely to increase production volume rather than reduce absolute emissions.

**Path Dependencies:** Across interviewees, AI is seen as a tool that functions to extend reserves and asset lifetimes, lowering costs and reinforcing carbon lock-in. Expert B reports that AI is applied "across the entire value chain of oil and gas" and is "used… to increase production volumes," which helps companies "remain competitive for longer." They draw a parallel to fracking, where new techniques unlocked "completely different resources than before," and warn that firms seek to "push prices down… and pull money out of the fossil fuel business for as long as possible." Lower prices meet only partly elastic demand, which means cost reductions slow switching to alternatives and reinforce lock-in.

On the supply side, Expert A explains that many basins remain underexplored due to technological limits and data gaps. AI can use historical patterns to "prognosticate… the exact or most feasible location where you can go and drill and get the well," and to anticipate "how the reservoir and how the subsurface is going to behave in the upcoming years." These capabilities guide investment and operational strategies, with the expectation that current production can be raised over time. Additionally, they extend the oil and gas supply by lowering the cost of exploration, which increases the likelihood of finding active wells.

Finally, Expert B notes a tightening coupling between AI demand and fossil supply. They point to the data center boom, where facilities "want to have a gas connection installed so that they can generate electricity on site using gas," and to AI developers that may become "major consumers of fossil fuels." This mutual reinforcement links digital expansion to gas-fired capacity, which further entrenches the fossil system and deepens path dependence.

**Infrastructure Integration:** The experts show mutual reinforcement between AI infrastructure and fossil systems, where AI both depends on and expands gas-fired capacity. The infrastructure coupling of AI into the oil and gas sector is discussed in two ways by the experts. Expert A provides an example of AI being integrated into critical infrastructure. They highlight a case where India's Oil and Natural Gas Corporation Limited "utilized AI and they tried to find out the exact location of the oil and gas reserves." After drilling the suggested sites, "the AI pinpointed three locations and all three wells were productive." This demonstrates how AI is becoming a functional component of the industry's capital-intensive exploration and production infrastructure.

Expert B describes a more systemic form of integration, where the infrastructure of the AI industry is becoming physically dependent on fossil fuels. They observe that "with the data center boom […] the first data centers […] want to get a direct gas connection […] to generate electricity directly on-site via gas." This trend positions data centers, which are essential for large-scale AI, as "major offtakers of fossil fuels." This creates a reinforcing feedback loop or "a nice entanglement […] almost like a cycle," where the expansion of AI infrastructure drives demand for the very fossil fuels that AI helps to extract more efficiently.

**Trade-offs & Competing Objectives:** AI systems, in the oil industry, are often optimized for a single objective, resulting in trade-offs that overlook environmental factors. Both experts detail how single-objective optimization for output sidelines environmental constraints during model design. Expert A gives a clear example. When AI is used to find new reserves, "it only puts the boundary conditions according to the oil and gas, but it does not consider other things like geology or maybe other parts of the environmental aspects". The example shows how optimizing AI to discover reserves can conflict with environmental concerns. If not considered during algorithm design, adverse environmental outcomes are likely. Expert B makes these competing objectives explicit, noting that "optimizing transport, it's all well and good, but it can also reinforce the underlying problem."

**Digital Solutionism & Framing:** Both experts note that the presentation of AI use in the oil and gas industry is shaped by branding choices and framing that emphasize technical progress while downplaying environmental implications. Expert B exemplifies this, noting that Microsoft "renamed their page from 'AI in Oil & Gas' to 'AI in Energy'" after public criticism. They further notice that the use of AI in "oil and gas [is] primarily portrayed positively" in the International Energy Agency's 2025 "Energy and AI" report. They add that public communication tends to keep a low profile in climate contexts. Expert B cautions that such framing "quickly leads to greenwashing," and urges scrutiny of whether a claimed efficiency gain is primarily a cost saving with a minor energy benefit as a byproduct.

### 4.5.3. Manifested Harms and Impacts

**Material Harms:** Material harms of AI-accelerated exploration and drilling are considered in both interviews. The experts note that the impacts of AI-accelerated exploration are increased physical harm to the environment and local communities. Expert A is unequivocal about the industry's imperative to drill once a discovery is made, stating that it "will be done, irrespective of environmental damages, irrespective of the people. Everybody will be relocated or whatever. This is how progress is started." This highlights a process where environmental harm and the relocation of communities are treated as externalities.

Expert B elaborates on the specific types of environmental impacts that occur as AI enables access to new, often more remote, locations. They state, "I can well imagine that you then have to access new places all the time. And yes, that surely has to do with logging and probably also with a leak here and there." These perspectives showcase how the entrenchment of AI in the oil and gas industry leads to material harms.

**Epistemic & Cultural Harms:** Expert A discusses how the deployment of AI in this context can lead to epistemic and cultural harms. Their statement that wells will be drilled irrespective of people and communities, thus implying forced relocation, presents a cultural harm.

This is compounded by an epistemic harm, where the narrow, optimized logic of AI systems can override more comprehensive human or local knowledge. Expert A notes that an AI tasked with finding resources "might not tell you all those things" related to environmental or geological context, concluding that "human intervention is required". The system, however, is structured to prioritize the AI's narrow findings, thereby devaluing the very human oversight needed to prevent adverse outcomes.

**System-Level Consequences:** At a systemic level, the experts synthesize how these dynamics entrench carbon lock-in, create market vulnerabilities, and heighten data and integrity risks. The primary system-level consequence is the reinforcement of carbon lock-in, which delays a global energy transition. Expert B's "main concern" is that AI allows companies to "remain competitive for longer [...] to pull money out of the fossil fuel business for as long as possible". Expert A concurs, stating that as AI accelerates extraction, it becomes "very difficult for us to become carbon neutral [...]". This lock-in is further strengthened by the previously discussed emerging feedback loop where AI data centers are becoming "major offtakers of fossil fuels," creating "a nice entanglement there." This dynamic means the growth of the AI sector can directly drive demand for fossil fuels, potentially creating a systemic codependency. Beyond entrenchment, the deep integration of AI into competitive market dynamics also creates new vulnerabilities. Expert A highlights that "a lot of time data risks are there," as companies attempt to acquire competitor data on bidding or production to "play with it". This introduces systemic risks of industrial espionage, market manipulation, and correlated failures.

### 4.6. CASE STUDY: RECYCLING & WASTE MANAGEMENT

For this case study, we consider the systemic environmental risks of AI when deployed in waste management. We base the case study on two expert interviews. We interviewed an academic robotics researcher (Expert A) and the machine learning lead from a major waste robotics company (Expert B). Both interviews outline that AI-powered robots are primarily used in the final quality control stage of waste sorting, where they supplement or replace humans, identifying and separating materials from mingled streams. In waste management, AI is framed as a technological solution to a complex environmental problem. Both experts conclude that it has the potential for systemic environmental risks. We apply our framework to illustrate how AI can both promise environmental benefits and potentially pose systemic environmental harm.

### 4.6.1. Structural Conditions

**Power Concentration & Asymmetries:** During both interviews, the experts highlighted how vendors centralize models and data, raising barriers to independent auditing and public interest research. Expert A comments on the concentration of market power, identifying the leading actors as "companies like AMP Robotics, Zen Robotics, or Waste Robotics. They further elaborate that the companies "own that data, they collect that data from all these facilities," highlighting how power is further concentrated in data ownership. Expert B supports that conclusion, commenting on their companies' data arrangement "we own our data recorded on their site." This centralized model is further noted by Expert A, explaining that recycling facilities that purchase the robots must then "pay a recurring fee to get their AI systems updated" with improved models. Expert A further notes how this ownership model increases barriers for independent research, as they do not have access to "millions and millions, maybe at this point billions of data points to train their algorithms".

Both experts also note that the concentration reflects global asymmetry in labor. As Expert B states, some competitors use a "traditional system where they send images from the waste feed for humans to annotate in some cheap-labor country". The practice of low-wage data labeling creates a global asymmetry where those labeling are unlikely to benefit from the profits or improved waste management.

**Governance:** This section identifies that safety rules cover hardware while AI accuracy is left to market rejection, creating a permissive environment for risk. Expert B provides insights into governance and regulation. They explain that the importance of regulation for the recycling industry as a whole, stating "Regulations have a big effect on our whole business, of course, because in different countries, there are different regulations of what needs to be recycled." However, they state that when it comes to AI's performance, there is "nothing" in terms of regulation. Existing rules focus on physical machine safety, such as requiring a "cage around the robot," but there is "no standardization there" to ensure the AI's sorting accuracy meets environmental permits. This regulatory gap fosters a system of self-regulation and market-based control. Performance monitoring is ultimately the "customer's responsibility" after the system is commissioned. The primary enforcement mechanism is the market itself. As Expert B explains, the "final control is the buyer of the material," who will "return a load of sorted material if it's not of the quality that they expect". Thus, in waste management, AI governance is characterized by industry-specific regulations and self-regulation concerning the AI models themselves. This regulatory gap creates a permissible environment for potential systemic environmental risks to arise.

**Market-Driven Imperatives:** Both experts consider economic incentives, rather than environmental objectives, as primary drivers of AI adoption in the waste industry. Expert B explains that a customer's decision is often based on a simple return on investment calculation: "So they just compare our solution to the pre-existing human solution, labor cost, and so on. And then if that payback time for the investment is less than two years or something, they move forward". They further note that the desire to "get rid of people" to avoid high labor turnover and potential liabilities from hazardous working conditions drives AI adoption. Expert B elaborates that waste management firms aim to "get some gate fee for the material" or "get rid of the material as cheaply as possible." Hence, it is in their economic interest to minimize the material that goes to burning. These dynamics illustrate how the waste management industry is primarily profit-oriented, with environmental benefits mainly due to regulation. Therefore, AI adoption is profit-driven.

Competitive pressures also play a role. Expert A notes that there is a "feeling of, 'Oh, are we falling behind? […] if I don't, will I get bankrupt?'" which can push facilities to purchase robotics without fully understanding their value. Expert A notes that pressure is intensified by the fact that profit margins in the industry are known to be "quite low". Expert B concurs that "there is also, and I think this is maybe even a larger group, which put extra value in having AI and robots compared to humans. Either because that gives them an edge in their own branding, or because they want to get rid of people." These dynamics show that AI adoption is primarily driven by profit-incentives and competitive dynamics rather than potential environmental benefit.

**Opacity:** Both strategic and inherent opacity obscure the performance and impact of AI sorting robots. The experts describe hype, proprietary data, and site-specific variability that obscure performance at deployment time. Expert A describes how a "big hype was created around these robots," with facility owners expecting "much more than what these robots could do". This hype is fueled by marketing from robotics companies, but customers "can only know their actual impact for your facility after you test them over a certain period of time".

The proprietary nature of the technology compounds this. The data and models are owned by the robotics companies, preventing independent auditing. Expert B explains that while they can offer a fire-walled "silo for such a customer" who demands data privacy, this comes at a cost, as the customer will not "get benefit from the others" who contribute to the shared, centrally trained model. This strategic opacity contributes to the unpredictability of the technology's real-world performance. Expert A states plainly that there is anecdotal evidence suggesting that the robots are currently "not delivering what people hoped for" and that recycling companies are "not happy with this robotics technology at this point".

4.6.2. Risk Amplification Mechanisms

**Rebound Effects:** While AI promises to make recycling more efficient, Expert B raises the concern of a rebound effect, articulating the "pessimist view" that such technological advancement "just increases the amount of production". In this scenario, improved recycling does not displace the use of virgin materials but supplements it, allowing for a net expansion of material consumption. As the expert frames it, "you anyways take all the virgin materials you can, and in addition, you do more production from the recycled materials".

**Path Dependencies:** The experts discuss how subscription models, vendor data control, and sunk costs lock facilities into potentially underperforming solutions. Once a recycling facility invests in a specific robotics platform, it enters a path-dependent relationship with a vendor. Expert A notes that the facilities may "stick to this longer" to justify their initial investment, even if performance is suboptimal. This technological lock-in is reinforced by the business model, where facilities must pay a "monthly fee or yearly fee so that your AI system gets updated with newer and newer models". The data ownership structure, where the vendor controls the data, further solidifies the dependency. Expert B concurs that their company follows this business model. Additionally, the sunk cost of investing in robots can be considered a form of path dependency. Finally, as both experts note, AI-investment can be driven by the desire to reduce human labor, further increasing the dependence on robots by reducing the workforce.

**Trade-offs & Competing Objectives:** Both experts discuss the explicit purity-throughput trade-off that can erode environmental outcomes. Expert B explains that this optimization is done "depending on the customer needs, case by case". Some customers want to maximize throughput, while others want to maximize purity. The AI can be tuned accordingly. For instance, "if you want to have a good purity, then you only pick the most confident things," but this necessarily lowers the recovery rate of recyclable materials. Expert A notes the same trade-off, stating, "if you are prioritizing throughput and don't have the required accuracy […] the product quality decreases [...] less environmental benefit [...] You might not be able to recover certain items at all." These competing objectives mean that environmental goals are balanced against commercial targets.

**Algorithmic Bias:** Algorithmic bias arises in AI-based waste stream management due to data variability. The experts identify this as a significant technical and environmental risk. They note that models often fail to generalize from one context to another and link domain shift in waste streams to model misclassification, with site-specific drift producing systemic performance failures.

Expert A explains that waste streams are in constant flux due to "new recyclables that come, others go all the time," as well as "seasonal changes" and regional differences in consumption. Their research provided direct evidence of this challenge. After collecting data from two different facilities in the United States, they found that a model "trained on one facility was not successful in the other facility, even though the geographical distance is not that much". This demonstrates that even with large amounts of data, adapting to local variations in waste composition and facility conditions like lighting remains a difficult problem.

The industry's response to this challenge is varied. Expert B describes how their company initially tried "to build one universal model that understands everything" but soon found the effort "very costly" and conceptually flawed. In response, they pivoted to a more agile strategy, making it "as easy as possible to change the models or adapt them and to build models from scratch". Their current process uses "active learning," where they "let the robot make mistakes and then correct those" to focus training on difficult or novel items, such as a "Christmas edition of Coca-Cola bottle" that a standard model might miss. If a new site is "quite different from what we expected," they can quickly "build reasonably good models for new sites," sometimes in as little as a week for one person.

However, this adaptive approach is not standard across the industry. Expert B explains that their competitors, who often use a slower and more expensive process of sending images to "some cheap-labor country" for annotation, have a "much harder time adapting". This leads them to "force this one-size model to fit all sites," which creates a systemic risk of deploying biased and underperforming models. This is a critical point of failure, as the success of AI in recycling depends on its ability to adapt to the specific and constantly changing nature of local waste streams.

**Digital Solutionism & Framing:** Expert A points to a techno-solutionist framing in the adoption of AI in waste management. They note that the logic behind single-stream recycling, where the idea was "if we throw technology at that, we might be able to figure out how to sort these," is flawed. Expert B considers this to be due to a significant gap between this promise and reality. A "big hype was created around these robots," but their performance has been disappointing. To be effective, sorting requires extremely high accuracy, but as they state, "we are not there".

### 4.6.3. Manifested Harms and Impacts

**Material Harms:** Material harms arise through the deployment of inadequate or mal-adapted AI systems in waste management. Expert A explains that if accuracy is sacrificed for throughput, the "product quality decreases, so you are actually getting less environmental benefit from this recycling process". Poorly sorted paper, for example, might be immediately "downgraded" to single-use napkins instead of being recycled into higher-quality paper, which could potentially require more virgin materials than if human sorters were used.

Expert B identifies a future risk where AI scanners are trusted without independent verification. They provide the example of sorting biowaste for compost. If a faulty AI scanner allows plastics into the final product, "it could be that you're putting plastics and whatnot to the sides of roads under the grass," and the contamination is only realized much later.

**Epistemic & Cultural Harms**: Both experts consider how the deployment of AI sorting systems can lead to the erosion of both local and contextual knowledge. Expert A emphasizes the value of human workers who possess local knowledge, as they "know their local trash". This nuanced understanding is difficult for a centralized AI model to replicate. Furthermore, Expert B describes their company's pragmatic solution to data bias. Instead of deeply investigating the reasons for site-specific differences, they "don't even necessarily try to understand very well why the task is now different. We just do a new one". This approach, while efficient, represents an epistemic shift away from deep understanding toward rapid, functional problem-solving driven by AI. Both examples highlight how AI poses systemic environmental risk by centralizing knowledge.

**System-Level Consequences:** At a systemic level, the experts synthesize how competition, opacity, and verification risks can normalize underperformance and enable regulatory evasion.

Expert A describes a fundamental system-level failure rooted in the industry's competitive dynamics. They explain that although "multiple companies competing with each other" are "trying to solve an environmental problem," they are "not sharing this information that could have been really useful to do so." This lack of collaboration is driven by fear that if they are "naive, they might get pushed out of the market," especially since "the profit margins are known to be quite low". This competitive pressure also leads to irrational technology adoption, where a company might buy robots "without necessarily knowing what their actual value is" out of a "feeling of, 'Oh, are we falling behind?'". This is compounded by the fact that the technology is currently underperforming.

Looking forward, Expert B identifies a significant potential systemic risk not in the sorting process itself, but in the verification systems that will follow. They explain that, as verification, "AI scanners of waste get more standard," there is a "big potential for a company to provide a waste scanner that says that the stuff that you put to burning is all residue […] independent of what the stuff is. Because that's what the plant owner wants to see". This creates a risk of systemic fraud to "escape the regulations" unless an entity "controls that the AI doesn't lie".

# 5. Distribution & Accountability

Across the AI life cycle, environmental burdens are disproportionately concentrated in resource-extractive and manufacturing regions, as well as among vulnerable socio-economic groups. Conversely, the environmental benefits of AI applications tend to localize in well-resourced regions. While accountability for these harms is often dispersed across complex global supply chains, the literature highlights responsibility among upstream suppliers, cloud infrastructure providers, model developers, and policymakers. Falk et al. (2024) find that the extraction of minerals, production of hardware, and disposal of electronic waste impose significant environmental and social harms on communities in the "Majority World" ("Global South"). In contrast, the advantages of AI development and use are concentrated in the "Minority World" ("Global North"). The AI Now Institute has compared the uneven regional distribution of AI's environmental costs to "historical practices of settler colonialism and racial capitalism", highlighting the severity of these disparities (Kak & Myers West, 2023). Additionally, UNESCO's (2021) recommendation on AI ethics cautions that "AI should not be used" if it creates "disproportionate negative impacts on the environment". In the following, we examine how environmental burdens and benefits are reported to be distributed across the AI supply chain and how responsibility for these externalities is often framed or displaced across regions.

The material burdens of the AI life cycle are borne by extraction sites and water-stressed regions, which are frequently located in LMICs and in disadvantaged areas in HICs (Domínguez Hernández et al., 2024; Falk et al., 2024). Over 70% of the world's cobalt, for instance, is produced in the Democratic Republic of Congo, often under poorly regulated conditions (Thomas et al., 2025). Similarly, the majority of REE refining is situated in China, such as Bayan Obo (Dauvergne, 2022; Falk et al., 2024). The mining activities have a "long history of environmental and human rights abuses" (Dauvergne, 2022; Domínguez Hernández et al., 2024; Falk et al., 2024). Only a negligible fraction of these critical materials is recycled, meaning primary extraction continues to drive ecological damage (Dauvergne, 2022). Notably, these activities occur in what Thomas et al. (2025) describe as "low governance areas," where enforcement of environmental and labor protections is minimal.

The burdens of the following manufacturing phase are concentrated in a relatively small number of regions that host semiconductor fabrication and electronics production facilities (Domínguez Hernández et al., 2024; Falk et al., 2024; Thomas et al., 2025). Falk et al. (2024) note that hardware manufacturing for AI is geographically concentrated in parts of East Asia, as well as in the United States and a few European countries. In Taiwan and South Korea semiconductor fabs have faced scrutiny for high water consumption and chemical waste, while in China and Southeast Asia, electronics assembly plants can generate pollution that affects local air quality and waterways (Falk et al., 2024). Li et al. (2024) observe that beyond energy use, "the environmental toll of chip manufacturing" includes toxic chemical exposure and waste as well as the upstream impacts of "raw material extraction".

Furthermore, the environmental impacts of data centers are unevenly distributed, often depending on a region's climate and energy infrastructure, resulting in varying emissions and water usage among data centers by location (Falk et al., 2024; Li et al., 2024). For instance, Water Usage Effectiveness (WUE) is highly sensitive to outside temperature (Li et al., 2024). In temperate or cooler climates, state-of-the-art data centers can achieve WUE below 1.0 L/kWh, using mainly air-cooling (Li et al., 2024). In contrast, under hot, arid conditions, the WUE can reach up to 9 L/kWh (Li et al., 2024). Such water extraction can exacerbate local water stress and drought conditions in already arid regions (Li et al., 2024). A similar regional disparity exists for carbon emissions. Li et al. (2024) report that as of 2020, only 4% of the energy used by Google's data center in Singapore was carbon-free, compared to 94% in its Finnish data center. These emissions also pose local issues, such as the emission of air pollutants from fossil-based electricity generation, which affects regional air quality (Li et al., 2024).

During the EOL stage of the life cycle, a significant portion of waste is exported from HICs to LMICs, where informal recyclers and nearby residents face toxic exposure from processing methods, as only approximately 20% of this e-waste is formally recycled worldwide (Falk et al., 2024). Studies document elevated levels of lead, mercury, and other toxins in soil and food near informal e-waste sites (Falk et al., 2024).

The distribution of negative and positive impact echoes global inequities: it has been characterized as a form of "algorithmic colonialism" (Nost & Colven, 2022), wherein resources and labor from LMICs sustain high-tech development that primarily serves HICs (Dauvergne, 2022; Domínguez Hernández et al., 2024; Falk et al., 2024; Thomas et al., 2025). This pattern also occurs within regions. Data centers are frequently located outside urban areas, which can require expansions of local utilities and infrastructure (Falk et al., 2024). Because AI inference workloads are routed across geographically distributed data centers to serve users in other regions, the locations that bear environmental burdens often differ from those that receive the benefits (Li et al., 2024). In practice, the fragmentation of AI creates an accountability gap, as each actor can claim that the most serious environmental impacts fall outside their responsibility (Falk et al., 2024). Addressing this gap would require moving beyond voluntary offsets and corporate branding (Dauvergne, 2022; Thomas et al., 2025). It calls into question the very design of these systems, pointing toward the need for governance models that embed equity directly, such as equity-aware geographical load balancing (Li et al., 2024), and robust assessment frameworks that mandate proportionality tests *before* deployment (Rohde et al., 2024; van Wynsberghe, 2021).

The transnational nature of AI services facilitates stakeholders' ability to displace or obscure accountability for negative externalities (Dauvergne, 2022; Domínguez Hernández et al., 2024; Thomas et al., 2025). As Dauvergne (2022) observes, the states and corporations that benefit most from AI tend to conceal or normalize the collateral damage. When environmental costs come to light, they are often framed as inevitable by-products of job creation, economic growth, and technological advancement (Dauvergne, 2022). However, these gains frequently do not translate to absolute reductions in environmental pressure (Dauvergne, 2022). Moreover, many large AI firms have adopted strategies such as purchasing carbon offsets or renewable energy credits to brand themselves as carbon-neutral (Falk et al., 2024; Thomas et al., 2025). Thomas et al. (2025) caution that treating specific impacts as simply offsetable externalities is insufficient, as they cannot compensate for other localized harms. In practice, the fragmentation of AI creates an accountability gap, as each actor can claim that the most serious environmental impacts fall outside their responsibility (Falk et al., 2024).

# 6. Conclusion

In this report, we synthesize a three-level framework of systemic environmental risks of AI, moving from structural conditions to risk amplification mechanisms and then to manifested impacts. Through the framework, we provide an overview and characterization of emergent risk pathways that may arise. Our analysis shows that AI does not produce static externalities but rather a dynamic, potentially self-reinforcing system of environmental risk. We illustrate this system with case studies aiming to ground our analysis in practice. We supplement our analysis of systemic environmental risks by reviewing the direct environmental impacts of AI, discussing potential trade-offs and mitigation strategies, and considering the distribution of AI's costs and benefits.

Our core finding is that AI's environmental risks are structural. They extend beyond energy and material footprints to reorganize decision-making and infrastructures in ways that accelerate environmental harm while diffusing accountability. These harms are unevenly distributed. Environmental burdens, from mineral extraction and water withdrawals to toxicity and e-waste, concentrate in resource extraction regions and in low and middle-income countries. At the same time, benefits accumulate in resource-rich actors. This pattern reproduces historical extractive relations. While AI shares many direct and systemic risks with the broader ICT sector, our analysis reveals that AI acts as both an accelerant of existing risks and an originator of novel systemic risks that legacy ICT governance models are ill-equipped to address. Without purposeful intervention, current trajectories position AI to accelerate rather than mitigate environmental harm.

We propose four policy interventions to address these dynamics:

- Mandated and standardized life cycle and systemic risk assessments, with proportionality tests that weigh social benefits against environmental and justice costs, should be prerequisites for model release, large-scale deployment, and major infrastructure expansion.
- Replace self-reporting with transparency and independent verification across data provenance, models, and direct and systemic environmental impacts, with public reporting of standardized metrics.
- Establish a precautionary global governance framework that treats AI as infrastructure with planetary-scale consequences and aligns development with planetary boundaries.
- Democratize data, models, and infrastructure and ensure that development and deployment are participatory.

We acknowledge the incompleteness of this analysis and that emergence, uncertainty, and opacity challenge the analysis. However, we consider these limitations to be a reason for precautionary governance and AI development. With AI increasingly becoming ubiquitous, decisions taken now may lock in trajectories for decades to come.

Our analysis supports a paradigm shift. AI development and use must move from growth-driven and opaque practices to transparent and precautionary governance. AI should primarily be considered as a planetary-scale infrastructure rather than an economic asset. As such, aligning its development and use with finite Earth systems should not be treated as an aspiration but rather as a near-term necessity.

THE SYSTEMIC ENVIRONMENTAL RISKS OF ARTFICIAL INTELLIGENCE    GESELLSCHAFT FÜR INFORMATIK   65Gröger, J., Behrens, F., Gailhofer, P., & Hilbert, I. (2025). Environmental impacts of artificial intelligence. Greenpeace. https://www.greenpeace.de/publikationen/20250514-greenpeace-studie-umweltauswirkungen-ki.pdf

Hacker, P. (2024). Sustainable AI regulation. Common Market Law Review, 61(2), 345–386. https://doi.org/10.54648/cola2024025

Haider, J., Rödl, M., & Joosse, S. (2022). Algorithmically embodied emissions: The environmental harm of everyday life information in digital culture. Proceedings of the 11th International Conference on Conceptions of Library and Information Science, 27. https://doi.org/10.47989/colis2224

Han, Z., Golev, A., & Edraki, M. (2021). A review of tungsten resources and potential extraction from mine waste. Minerals, 11(7), 701. https://doi.org/10.3390/min11070701

Heilinger, J.-C., Kempt, H., & Nagel, S. (2024). Beware of sustainable AI! Uses and abuses of a worthy goal. AI and Ethics, 4(2), 201–212. https://doi.org/10.1007/s43681-023-00259-8

Hendrycks, D., Mazeika, M., & Woodside, T. (2023). An overview of catastrophic AI risks (No. arXiv:2306.12001). arXiv. https://doi.org/10.48550/arXiv.2306.12001

Hess, J. C. (2024). Chip production's ecological footprint: Mapping climate and environmental impact. Interface. https://www.interface-eu.org/publications/chip-productions-ecological-footprint#

Heydari, A., Gharaibeh, A. R., Tradat, M., soud, Q., Manaserh, Y., Radmard, V., Eslami, B., Rodriguez, J., & Sammakia, B. (2024). Experimental evaluation of direct-to-chip cold plate liquid cooling for high-heat-density data centers. Applied Thermal Engineering, 239, 122122. https://doi.org/10.1016/j.applthermaleng.2023.122122

Hoffmann, L., Zeck, T., & Becker, N. (2025). Auswirkungen von KI, Rechenzentren und Halbleitern auf Wasserverfügbarkeit und -Qualität. Gesellschaft für Informatik. 10.18420/studie_KI2025_01

Huang, Y.-C., Lin, Y.-X., Hsiung, C.-K., Hung, T.-H., & Chen, K.-N. (2023). Cu-based thermocompression bonding and cu/dielectric hybrid bonding for three-dimensional integrated circuits (3D ICs) application. Nanomaterials, 13(17), 2490. https://doi.org/10.3390/nano13172490

IEA. (2025). Energy and AI. International Energy Agency. www.iea.org/reports/energy-and-ai

IEEE Electronics Packaging Society. (2024). Chapter 10: Integrated Power. In Heterogeneous Integration Roadmap (2024 Edition). IEEE Electronics Packaging Society. https://eps.ieee.org/technology/heterogeneous-integration-roadmap.html

# GESELLSCHAFT FÜR
# INFORMATIK E. V. (GI)

**Office Bonn**
Ahrstr. 45
53175 Bonn
Phone: +49 228 302-145
Fax: +49 228 302-167
E-Mail: bonn@gi.de

**Office Berlin**
Weydingerstr. 14–16
10178 Berlin
Phone: +49 30 7261 566-15
Fax: +49 30 7261 566-19
E-Mail: berlin@gi.de

gs@gi.de
www.gi.de

/mas.to/@informatik
/company/gesellschaft-fuer-informatik